\documentclass[fleqn,usenatbib]{mnras}

\usepackage[T1]{fontenc}
\usepackage{ae,aecompl}

\usepackage{graphicx}	
\usepackage{amsmath}	
\usepackage{amssymb}	

\usepackage{listings} 
\usepackage{verbatimbox} 

\usepackage{todonotes} 
\usepackage{setspace}
\newcommand{\img}[3]{
\begin{figure*}
	\centering
	\includegraphics[width=#1\textwidth]{#2}
	\caption{#3}
	\label{#2}
\end{figure*}
}

\newcommand{\imgsinglecol}[3]{
\begin{figure}
	\includegraphics[width=#1\columnwidth]{#2}
	\caption{#3}
	\label{#2}
\end{figure}
}

\newcommand{\Lya}{Ly$\alpha$}
\newcommand{\Lyb}{Ly$\beta$}
\newcommand{\A}{\AA}

\newcommand{\mnhi}{N_{\rm HI}}
\newcommand{\nhi}{$\mnhi$}
\newcommand{\mdnhi}{\Delta \mnhi}

\newcommand{\mzabs}{z_{\rm abs}}
\newcommand{\zabs}{$\mzabs$}
\newcommand{\mzem}{z_{\rm em}}
\newcommand{\zem}{$\mzem$}
\newcommand{\confp}{{\it confidence}}
\def\cm#1{\, {\rm cm^{#1}}}

\newcommand{\nsdss}{4,913}  
\newcommand{\newsdss}{1,659} 
\newcommand{\highsdss}{2,714}  
\newcommand{\nboss}{50,969}  
\newcommand{\highboss}{18,959}  
\newcommand{\newboss}{9,230}  
\newcommand{\bossmatch}{10,571}  

\title{Deep Learning of Quasar Spectra to Discover
and Characterize Damped \Lya\ Systems} 

\author[Parks et al.]{%
David Parks,$^{1}$
J. Xavier Prochaska,$^{2}$
Shawfeng Dong,$^{3}$
Zheng Cai,$^{2,4}$
\\
$^{1}$Computer Science, UC Santa Cruz, 1156 High St., Santa Cruz, CA 95064 USA --- dfparks@ucsc.edu\\
$^{2}$Astronomy \& Astrophysics, UC Santa Cruz, 1156 High St., Santa Cruz, CA 95064 USA \\
$^{3}$Applied Mathematics \& Statistics, UC Santa Cruz, 1156 High St., Santa Cruz, CA 95064 USA\\
$^{4}$Hubble Fellow 
}

\date{Accepted XXX. Received YYY; in original form ZZZ}

\pubyear{2017}

\begin{document}
\label{firstpage}
\pagerange{\pageref{firstpage}--\pageref{lastpage}}
\maketitle

\begin{abstract}
We have designed, developed, and applied a convolutional
neural network (CNN) architecture using multi-task
learning to search for and characterize strong HI \Lya\
absorption in quasar spectra.    Without any explicit 
modeling of the quasar continuum nor application of
the predicted line-profile for \Lya\ from quantum mechanics, 
our algorithm predicts 
the presence of strong HI absorption and estimates the corresponding
redshift \zabs\ and HI column density \nhi, with emphasis
on damped \Lya\ systems (DLAs, absorbers with 
$\mnhi \ge 2 \times 10^{20} \cm{-2}$).
We tuned the CNN model using a custom training set
of DLAs injected into DLA-free quasar spectra from the
Sloan Digital Sky Survey (SDSS), data release 5 (DR5).
Testing on a held-back validation set demonstrates a high incidence
of DLAs recovered by the algorithm
(97.4\%\ as DLAs and 99\%\ as an HI absorber 
with $\mnhi > 10^{19.5} \cm{-2}$)
and excellent estimates for \zabs\ and \nhi.
Similar results are obtained against a 
human-generated survey of the SDSS DR5 dataset.
The algorithm yields a low incidence of false positives
and negatives but is challenged by 
overlapping DLAs and/or very high \nhi\ systems.
We have applied this CNN model to the quasar spectra
of SDSS-DR7 and the Baryonic Oscillation Spectroscopic
Survey (BOSS, data release 12) and provide catalogs
of \nsdss\ and \nboss\ DLAs respectively
(including \newsdss\ and \newboss\ high-confidence
DLAs that were previously unpublished).
This work validates the application of deep
learning techniques to astronomical spectra for both
classification and quantitative measurements.
\end{abstract}


\section{Introduction}
\label{section:introduction}

The dominant reservoirs of atomic hydrogen in the universe are traced
by layers of HI gas referred to as the damped \Lya\ systems (DLAs).
The DLAs are defined to have integrated HI column densities 
$\mnhi \ge 2 \times 10^{20} \cm{-2}$, sufficiently high to show
damping wings driven by quantum mechanic broadening of 
the HI \Lya\ transition.
These absorption lines are observed
in the spectra of distance sources that emit at far ultraviolet
wavelengths from gas both intrinsic to the source
and intervening.

\cite{wolfe1986} were the first to recognize that such absorption systems
could be surveyed in the distant universe through spectroscopy of
high $z$ quasars.  Their primary motivation was to reveal conditions
in the interstellar medium (ISM) of distant galaxies, expecting that
very high \nhi\ absorption would trace gas primarily within galaxies.
Using the 4m-class telescopes of that time, their initial
and subsequent surveys derived a sample of $\sim 100$ DLAs at $z \sim 2$
\citep{lanzetta,wolfe95}.  The discovery and analysis of these DLAs
were pursued primarily with 
visual inspection and by-eye fits to the \Lya\ absorption
using Voigt profiles.
These surveys established that DLAs are uncommon 
($\approx 1$ system per 4 sightlines) 
and offered the first estimates of the cosmological
mass density of neutral hydrogen.


Follow-up observations of metal-line transitions from heavy
elements enriching the gas have enabled additional scientific
exploration on the physical conditions within the interstellar
medium of distant galaxies.
This includes the chemical evolution of galaxies in 
the early Universe \citep{pettini1997dust,pgw00}, 
dust depletion, nucleosynthetic enrichment, 
and ionization of these protogalaxies 
\citep[e.g.][]{prochaska2002ucsda}, and constraints
on the dynamics of protogalaxies \citep{PW97}.
The gaseous clouds traced by DLAs also play an 
important role in fueling the star formation of galaxies 
at high redshift \citep{ota2014alma,rudie2017unique,neeleman2017cii}. 
Researchers have even identified novel applications of DLAs
for complimentary research, e.g. the
detection of QSO host galaxies using DLAs as 
natural coronagraphs \citep[e.g.][]{hennawi09,finley13,cai14}. 

On the other hand, DLA absorption occasionally compromises
the scientific analysis of quasar spectra.  This includes
measurements of the mean opacity and flux probability distribution
function of the intergalactic medium \citep[e.g.][]{lee15}, 
measurements of baryonic acoustic oscillations with the
\Lya\ forest \citep{slosar},
and searches for extraordinary overdensities (protoclusters)
via large effective opacities \citep{cai}.
The identification of bona fide DLAs is critical to 
pruning or masking undesired spectra in such analyses.

In 2003, the Sloan Digital Sky Survey (SDSS) published its first data release
\citep[DR1;][]{SDSS_DR1}
including $\approx 2,500$ 
$z>2$ quasar spectra.
This inspired a new generation of 
DLA surveys using these large datasets \citep{ph04}.  
When the project completed  with its seventh data release (DR7), 
the dataset included over 20,000 quasars at $z>2$ and several teams 
had developed a combination of human and algorithm-based approaches 
to survey the spectra for DLAs \citep{Prochaska2005,Prochaska2009,noterdaeme2009}.
With a sample of over 1,000 DLAs, the statistical uncertainties
on measurements of the incidence of DLAs and the shape of their
\nhi\ distribution function 
achieved $\sim 10\%$ precision or better.
This trend of increasing datasets and their greater statistical power
implies several challenges to future progress:
(i) even with custom tools, the human vetting process is expensive, prone
to error, and nearly irreproducible.  
(ii) humans may introduce systematic error 
comparable to or even exceeding
the statistical uncertainties.
In particular, standard techniques include a human-estimated
continuum to normalize quasar emission when analyzing
the DLA profile.

These issues inspired our group, and another team
\citep{Garnett2016}, to employ machine learning techniques
to generate fully automated analysis of quasar spectra for 
the survey of DLAs.  The \citet{Garnett2016} approach used
Gaussian processes to train
on data from the 9th data release of the BOSS survey \citep{Noterdaeme2012} 
to learn a model of the quasar emission spectrum without DLAs.
Their algorithm then estimates
the probability of a DLA occurring within a given 
spectrum and generates the probability distribution function
for its redshift and column density.
The authors applied this analysis to
quasar spectra in
the 12th data release of the BOSS
and reported 15,250 sightlines with probability $P > 0.9$
of at least one absorber with
$\mnhi > 2 \times 10^{20} \cm{-2}$. The authors further derived and reported point estimates of DLA parameters ($z_{\textsc{dla}}$, $\mnhi$) for presumed DLAs, using the \textit{maximum a posteriori} (MAP) estimation technique. Compared with the reported values in the DR9 catalog, their MAP estimates of the absorber redshift $z_{\textsc{dla}}$ are remarkably close to the catalog figures. However, their MAP estimates of column density $\mnhi$ show a lot more variation with the catalog figures. They thus recommended using the standard Voigt-profile fitting procedures to derive a more accurate estimate 
of the parameters, when desired. 
These authors have since applied results from their algorithm
to examine statistics of DLAs at $z>2$ \cite{bird17}.

Our approach  is motivated by the history of visually scanning quasar
spectra to discover DLAs, i.e.\ their signature 
is easily recognized by a well-trained astronomer.
Therefore, we 
created a convolutional neural network CNN)
model which is capable of identifying any number of DLAs in a sightline, 
and provide measurements of their absorption redshifts
and HI column densities. 
Convolutional neural networks are very effective in 
image processing and classification tasks. 
In the following, 
we treat the SDSS sightlines as one dimensional images,
apply standard CNN techniques, and innovate several
new approaches.

In processing a sightline we examine a sliding window
of 400 pixel segments, and ask the model to predict 3 results 
for each segment: 1) a classification 
of the segment as containing a DLA or not, 
2) the location of the center of the DLA in pixels 
(or 0 if there is no visible DLA), and 
3) the logarithmic column density if a DLA is visible.  
In our training, we exclude sightlines with known
BALs (\textit{broad absorption lines}) and do not attempt 
to classify them, though the model can 
conceptually be extended to include classification of BALs as 
a follow-up project.


This paper describes the development of a CNN architecture
designed to detect and characterize DLAs in the quasar spectra
of the SDSS and BOSS surveys.  We strive to develop an algorithm
that identifies DLAs at low redshift and in low signal to noise conditions, 
to handle DLAs near \Lya\ emission, 
to identify multiple and 
overlapping DLAs in the sightline, and to provide 
accurate column density measurements.  In addition,
we aim to achieve this without explicitly normalizing the
quasar continuum and without introducing any 
quantum mechanics to describe the line-profile of \Lya\
(i.e.\ no Voigt profile).
In fact, the algorithm even ignores the estimated variance
in the data (i.e.\ the recorded error array).
Instead, the CNN algorithm must `learn' these aspects on 
a large, custom training set.  

This paper is organized as follows.  
Section~\ref{section:terminology} provides a brief
description of the terminology related to surveys of
DLAs including the basics of quasar spectroscopy.
The neural network model is introduced in 
Section~\ref{section:nndetails} and our approach
to training the model is discussed in Section~\ref{sec:training}.
Validation of the model is performed in Section~\ref{section:validation}
where we examine results for a held-back sample from the 
training set and a human-generated survey of SDSS DR5
\citep{Prochaska2009}.
We present DLA catalogs derived from the SDSS-DR7 and
BOSS-DR12 in Sections~\ref{sec:dr7}
and \ref{section:dr12} and finish with concluding
remarks in Section~\ref{section:conclude}.
All of the code related to this project is available
at https://github.com/davidparks21/qso\_lya\_detection\_pipeline
together with complete catalogs of results.

\section{Basic Terminology of a DLA Survey}
\label{section:terminology}


In this section, we introduce the basic concepts of 
surveying DLAs in quasar spectra.  Experts in this field may
wish to skip to the following section.  Non-experts
are also referred to the review article on DLAs by \cite{Wolfe2005}.

By definition, a damped \Lya\ system is an absorption 
system discovered by its HI \Lya\ profile, which has
HI column density $\mnhi > 2 \times 10^{20} \cm{-2}$.
The gas is localized in redshift, although not with a rigorous
definition. The shape of the DLA 
follows a Voigt profile, derived from
Quantum Mechanic damping that motivates the DLA name. 
The wings span many hundreds and even
thousands of km/s, such that one cannot usually
resolve individual
DLAs with separations of less than $\approx 500$ km/s.
A reasonable working definition, therefore, is that the
integrated \nhi\ is over all gas within a 
$\Delta v = 1000$~km/s interval.
Further, the metal-line transitions observed
coincident with DLAs have widths that rarely
exceed a few hundred km/s \citep{PW97,Neeleman13},
i.e.\ these are well-localized within this $\Delta v$ definition.

\begin{figure*}
\includegraphics[width=\textwidth]{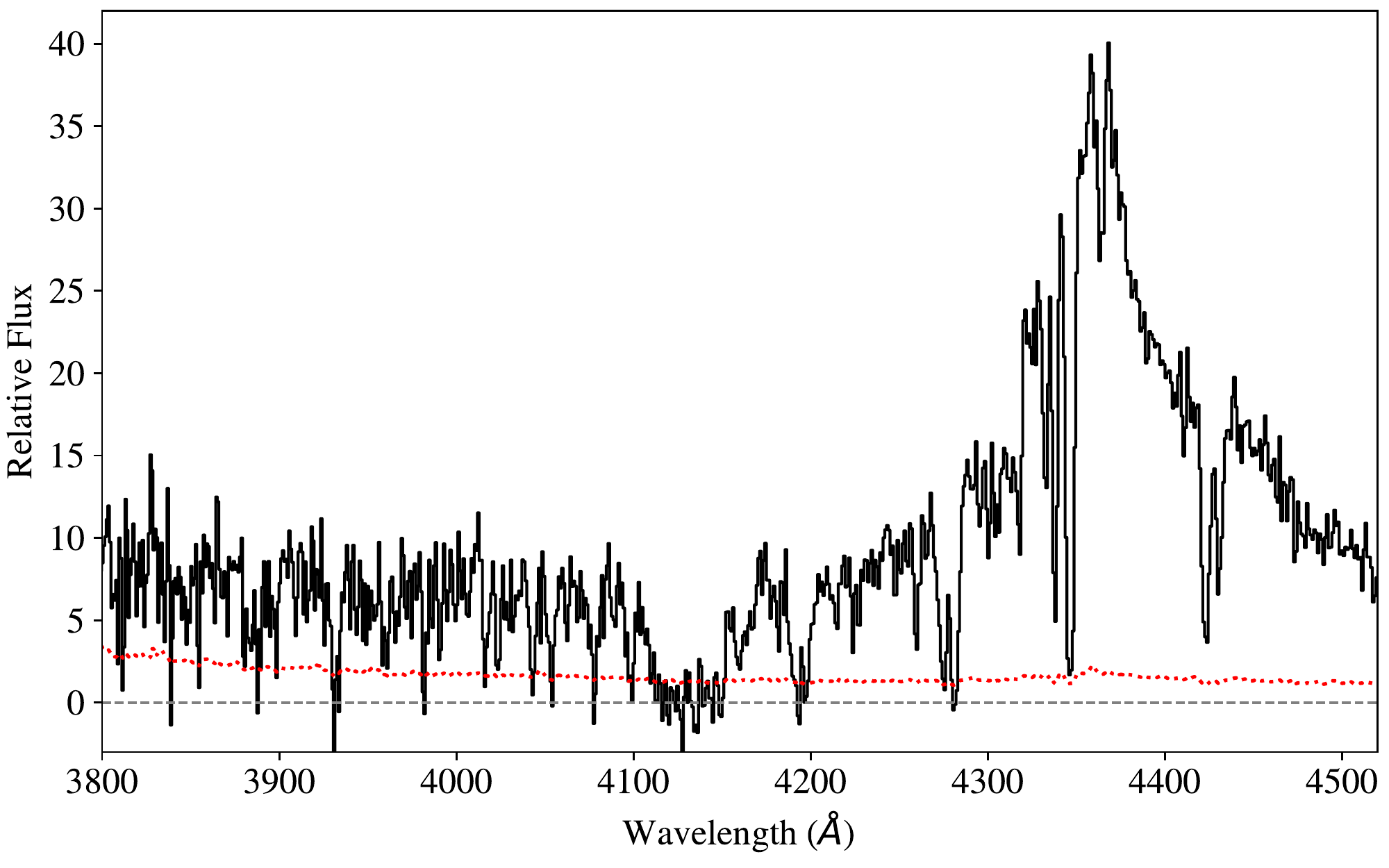}
\caption{
SDSS spectrum of quasar J150321.87+323742.2 at
$\mzem = 2.592$ which exhibits an intervening DLA
at $\mzabs \approx 2.397$ with 
$\mnhi \approx 10^{20.75} \cm{-2}$
\citep{Prochaska2005}, which can be seen on the graph at the 
wavelength $\lambda_{\rm obs} \approx 4120$~\AA.
}
\label{fig_dla_example.pdf}
\end{figure*}

Figure~\ref{fig_dla_example.pdf} shows the SDSS spectrum of a DLA
towards quasar  J150321.87+323742.2 
with emission redshift  $\mzem = 2.592$. 
These data have a spectral resolution FWHM~$\approx 150$\,km/s
and a dispersion of $\delta\lambda \approx 1$\AA\ per pixel.
The broad and strong \Lya\ emission of the quasar, emitted
at rest wavelength $\lambda_{\rm rest} = 1215.67$\AA\ in 
the quasar's rest-frame, is apparent at an 
observed wavelength $\lambda_{\rm obs} = \lambda_{\rm rest} (1+\mzem)
\approx 4380$\AA. 
The DLA occurs within the so-called {\it \Lya\ forest},
the thicket of absorption features blueward of the quasar's \Lya\ emission
due to hydrogen gas in the intervening, intergalactic medium 
at $z < \mzem$ \citep[e.g.][]{mcquinn16}. 
The DLA absorption is centered at $\lambda_{\rm obs} \approx 4130$\AA\
giving an absorption redshift 
$\mzabs = \lambda_{\rm obs}/\lambda_{\rm Ly\alpha}-1 \approx 2.397$.
Although not covered in this spectrum, the DLA also exhibits higher
order Lyman series lines including strong \Lyb\ absorption.

Traditional analysis of a DLA is to first estimate the unabsorbed quasar
flux near \zabs, referred to as the quasar {\it continuum}.  One then
estimates \zabs\ and \nhi\ for the DLA by comparing a model \Lya\ Voigt
profile, which is the convolution 
of a Lorentz profile predicted by Quantum mechanics in the dipole 
approximation\footnote{See 
\cite{lee03} for the next term from a more accurate
Quantum Mechanic treatment.} 
and a Gaussian profile assuming the gas 
satisfies a Maxwellian velocity distribution, to the absorption in the data.
Errors in the \nhi\ value are estimated from the quality of the
data, blending with the \Lya\ forest, and uncertainty
in the continuum estimation.

A survey for DLAs repeats this analysis for many
quasar spectra taking several considerations into account.
First, one typically sets a minimum threshold on the
{\it signal-to-noise} (S/N) of the spectra to insure
confidence in the detection and analysis of the DLA profile.
Previous human analyses of the SDSS quasar spectra
have typically restricted to S/N~$>4$, defined from the
median flux of the data to the reported error array.
Second, approximately 15\%\ of luminous quasars exhibit
strong associated absorption manifesting as 
{\it broad absorption lines} (BALs) in transitions
of highly ionized species (e.g.\ OVI, NV).  These BALs
mimic DLAs and tend to compromise the spectrum for a
DLA analysis.  As such, they are generally ignored
in DLA surveys.  

Previous DLA surveys \citep[e.g.][]{wolfe95,Prochaska2005}
have demonstrated that the incidence
of DLAs is modest, e.g.\ $\approx 1$ in 4 SDSS spectra
of $z>2.5$ quasars exhibit a DLA.
One also finds that the distribution of \nhi\ values
follows an $\alpha \approx -2$ power-law from column
densities $\log \mnhi \approx 20.3 - 21.7$, which then
steepens considerably at higher \nhi.

%

Absorption systems with \nhi\ just below
the DLA definition, i.e.\ $\mnhi \approx 10^{19.5-20.3} \cm{-2}$,
are commonly referred to as super Lyman limit
systems (SLLSs) or sub-DLAs \cite{prochaska2015}.
Uncertainty in the \nhi\ analysis implies a 
non-negligible number of SLLSs will be mis-classified
as DLAs (by any approach).  At low \nhi, these
are the primary contaminants in a DLA survey.



\img{}{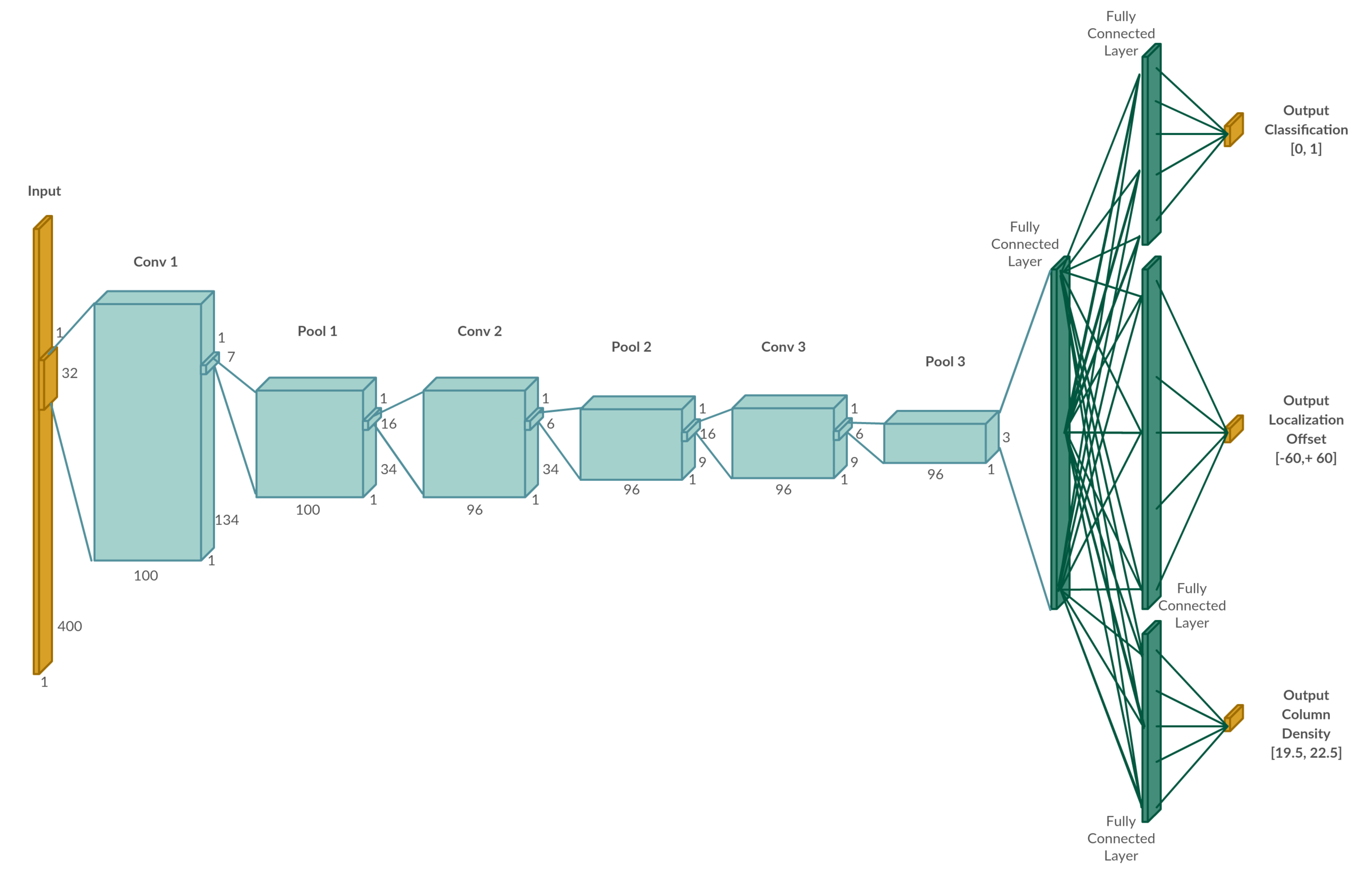}{An outline of the neural network architecture used. See section \ref{section:nndetails} for a more detailed description. This is a standard convolutional neural network most commonly applied to image problems. We essentially cast DLAs as a 1D image problem.}

\section{Neural Network Model}
\label{section:nndetails}

In this section we describe our neural network model. 
We use a standard 2D convolutional neural network architecture, but because a spectrum's flux is a one-dimensional dataset, one dimension of the CNN is simply fixed at 1. An overview of the overall architecture is visualized in 965Figure~\ref{neuralnetworkarchitecture.pdf}.

\subsection{Convolutional neural network architecture} 
\label{section:convolutional_network_architecture}

Convolutional neural networks exploit spatial locality by enforcing a local connectivity pattern between neurons of adjacent layers. The architecture thus ensures that the learned {\it filters} produce the strongest response to a spatially local input pattern. Stacking many such layers leads to nonlinear {\it filters} that become increasingly {\it global} (i.e., responsive to a larger region of pixel space). This allows the network to first create representations of small parts of the input, then from them assemble representations of larger areas. In our case, this process allows the neural network to learn high level concepts such as measuring column density, rather than simply memorizing the input. As such, convolutional networks generalize to unseen samples very well.

Our neural network is constructed using a fairly standard CNN architecture \citep{lecun-89t}. First, three convolutional layers are constructed, each with a max pooling layer. A fully connected layer of size 350 follows the convolutional layers. Then follows the last layer that is comprised of 3 separate fully connected sub-layers, the outputs of which are used to perform 1) classification, 2) DLA localization, and 3) column density estimate, respectively. This last layer of {\it multi-task learning} is a novelty of our model 
(section \ref{section:multi-task_learning}). 

The initial architecture started with classification only and evolved into three separate convolutional networks for the three outputs, each having hyperparameters tuned to achieve the best possible results. In experimentation we found that one network performed better on all three tasks than three individual networks. 
Our choice of three convolutional layers was initially arbitrary but experimenting with more layers did not improve results. 
All other hyperparameters were extensively optimized for, as discussed in section \ref{section:hyperparameter_selection}. The hyperparameters that govern the architecture and training of the network are tabulated in the Appendix (table \ref{-table:hyperparameters}).

We use the Adam (\textit{Adaptive Moment Estimation}) algorithm to search for the optimal parameters for our neural network model \citep{Adam2015}. Our neural network is implemented using Google's deep learning framework TensorFlow \citep{tensorflow2015-whitepaper}; and was trained on a server with two NVIDIA Tesla K20 GPUs.

\subsection{Data input} 
The input to the neural network consists of the raw flux 
values in the sightlines. 
The uncertainty values (i.e.\ error array) 
are entirely ignored by the algorithm.
Furthermore,  no scaling was applied to the flux data.
In fact, any attempt to perform typical zero-mean/unit-variance standardization, 
or to scale flux values to $(0,1)$ produced worse results 
than leaving them unscaled. 
Input values typically range from $\approx 0 - 50$
in the SDSS flux units of $10^{-17} \rm erg \, s^{-1} \, cm^{-2} \, \AA$. 

We analyze a fixed range of the sightline ranging from 900\AA\ to 
1346\AA\ in the quasar rest frame. 
The lower bound corresponds approximately to the HI Lyman limit in the
rest-frame of the quasar.  Below these wavelengths, intervening optically
thick HI gas may greatly absorb the quasar and partially mimic the 
signal of a DLA.  The upper bound of our analysis interval
scans past the \Lya\ line to ensure 
DLAs on or near the \Lya\ emission line are recovered\footnote{Most of these
$z > z_{\rm em}$ candidates, however, are instead strong NV absorption associated
to the quasar.}. 
From SDSS or BOSS spectra, this yields an input that is of 
size $[1748, 1]$ (padded as necessary; see below), 
which are the dimensions fed to the neural network. 
Our first layer convolutional kernel is then $[32,1]$. 
This kernel size is much larger than the typical $[3,3]$ or $[5,5]$ kernel 
seen in most vision problems\footnote{Well known vison architectures MobileNet \cite{Howard2017} and Inception V4 \cite{Szegedy2017} both use a $[3,3]$ kernel.}; 
we discuss how we chose this value 
and the hyperparameter selection process in detail in section \ref{section:hyperparameter_selection}.

We do not actually input all 1748 sightline pixels into our network because 
this would make identifying multiple DLAs along a given sightline too 
challenging\footnote{ 
For that approach, one would need to construct a training set that accounted
for all DLA configurations within the full span of these sightlines, a process
we considered currently intractable.}.
Therefore, the input to the network is 
a 400 pixel region of the sightline in a sliding window. 
For each sightline, the model performs 1748 separate inference steps
with a 400 pixel window around every pixel of the spectrum 
producing a result. 
See the visualization aid in 
Figure \ref{sliding_window_visual.pdf}. 

The choice of 400 pixels was dictated by the width of DLA absorption
($\approx 10$\AA\ in the DLA rest-frame) and the dispersion $\delta\lambda$
of the spectrometer ($\approx 1$\AA\ for the SDSS and BOSS surveys).
That is, we desired a window large enough to easily encompass a single
DLA without greatly exceeding its width.
Spectrometers with higher/lower dispersion would require a larger/smaller
window.  This means that the specific model described here
is not directly applicable to datasets with different
$\delta \lambda$, but we expect 
it will be straightforward to construct new 
models within a similar CNN architecture. 

\img{}{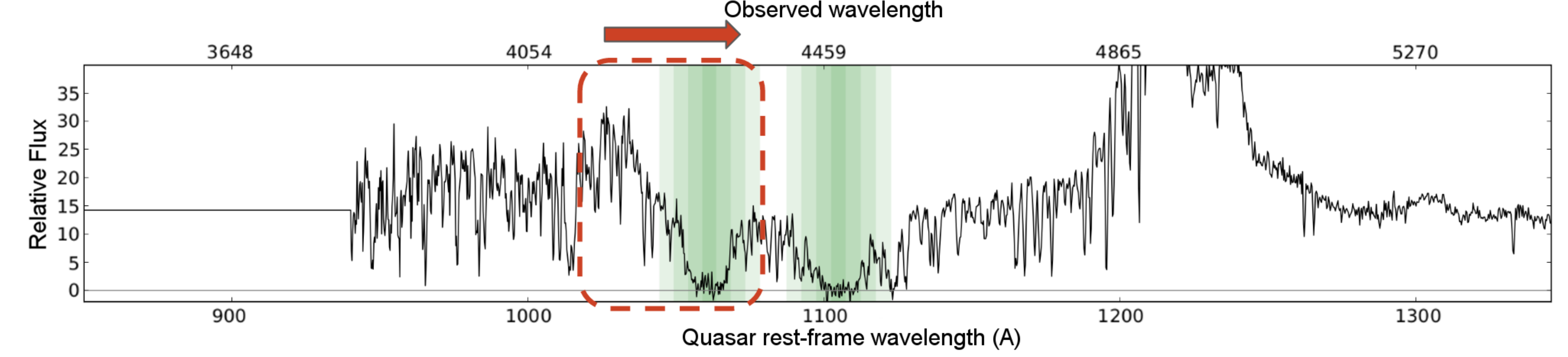}{The sightline is broken into 
400 pixel segments in a sliding window. 
An independent segment generates one prediction at
each of the 1748 pixels in the sightline covering
$900-1346$\AA\ as the center point in the quasar rest-frame. 
This approach facilitates identifying overlapping DLAs and 
also generates a large training dataset.}

\subsection{Multi-task learning} 
\label{section:multi-task_learning}
The network produces three outputs (labels)
for every 400 pixel window in the spectrum.
This multi-label approach
is a novelty of the model and not a typical application of CNNs. 
The algorithm first
classifies whether a DLA exists within 60 pixels of the center
of a 400 pixel window. The value 60 was chosen as a range for which we can safely assume the DLA will be fully visible in the 400 pixel window.
If a DLA is predicted the label takes on a 1/true value, else 0/false. 
Second the algorithm
estimates the DLA line-center by producing a value in the range 
$[-60,+60]$ that indicates how many pixels the window center is from the center of the DLA. 
Third, the algorithm produces a column density estimate that is only
valid if a DLA actually exists in the current window. 
A visualization of the labels for each pixel in a 
sightline is shown in Figure~\ref{labeling_diagram.pdf}. 

Our initial approach to the problem began with a simple CNN
that identified sightlines with or without a DLA (simple binary classification). We then created a separate network that performed localization by improving upon the simple binary classification and outputting a predicted center point of the DLA relative to the current input. Finally we trained a third model to take the samples where a DLA exists somewhere in the window and trained it to predict the column density of the DLA, regardless of its location. 
We then found that combining the models into one multi-task learning
model produced better results than training each model independently. 
Multitask learning refers to the concept that a 
learning algorithm, such as a neural network, may perform better when it is generalized to solve multiple (related) tasks. This is a method of regularization discussed in the Deep Learning Book by \cite{Goodfellow-et-al-2016}. 
With multi-task learning,
we noticed an improvement in all metrics over the individually 
specialized models.

An astute reader will note that using a single model for all 3 of these outputs 
will force us to train on areas of sightlines where no DLAs exist. In these regions,
the predictions for the column density measurement are irrelevant. 
We sidestep this issue by masking the gradient appropriately, as
described in section~\ref{section:maskingthegradient}.


\img{}{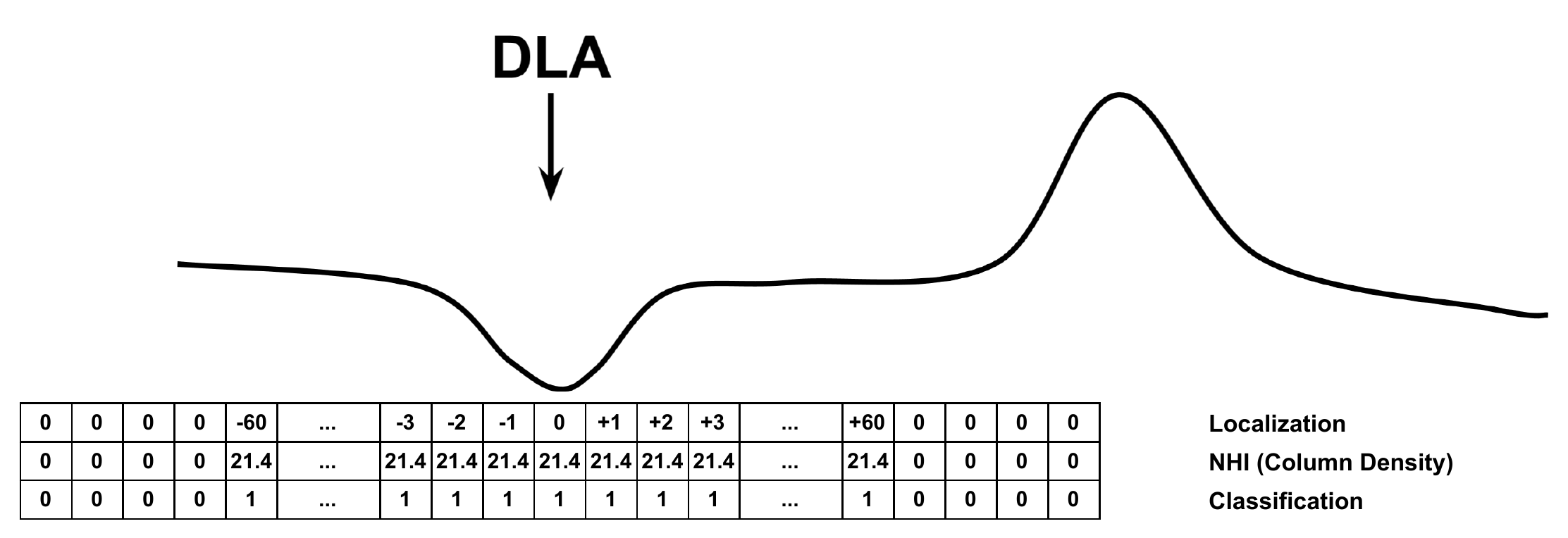}{A visual aid demonstrating the 
idealized labels for a DLA (not to scale). For any specific point taken 
as the center point of a 400\,pixel
window, the labels indicate: 1) the location (offset) of a DLA within that window or zero when no DLA exists, 2) the column density if a DLA exists in the window, or zero, and 3) a 1/0 classification indicating that a DLA exists.}

\subsection{Loss function for multi-task learning}
\label{section:cnn_loss}
In multi-task learning a model will produce multiple outputs for a given input. 
In our model, three outputs are produced: classification, localization, and column density estimation.
Here we document the three individual loss functions, and the combined loss function which is optimized for. See figure \ref{labeling_diagram.pdf} for a visual guide to the three labels we work with in these equations.

The loss function for DLA classification ($\mathcal{L}_{c}$) in a sample 
is the standard cross-entropy loss function:
\begin{equation}
\mathcal{L}_{c} = -y_c \log(\hat{y}_c) - (1-y_c) \log(1-\hat{y}_c)
\label{eqn:classification_loss}
\end{equation}
where $y_c$ is the ground truth classification label ($y_c = 1$ if DLA; 0 otherwise) for the sample; and $\hat{y}_c$ is the model's predicted classification ($\hat{y}_c$ is in the range (0,1) where $\hat{y}_c > 0.5$ indicates a positive DLA classification and $\hat{y}_c < 0.5$ negative classification).

The loss function for DLA localization ($\mathcal{L}_{o}$) is the 
standard square error loss function:
\begin{equation}
\mathcal{L}_{o} = (y_{o} - \hat{y}_{o})^2
\label{eqn:offset_loss}
\end{equation}
where $y_o$ is the ground truth localization label in the range $(-60, +60)$; 
and $\hat{y}_o$ is the model predicted localization.

The loss function for log column density estimation ($\mathcal{L}_{h}$) is a square error loss function with gradient mask (see section \ref{section:maskingthegradient}):
\begin{equation}
\mathcal{L}_{h} = \left(\frac{y_c}{y_c + \epsilon}\right)(y_{h} - \hat{y}_{h})^2
\label{eqn:columndensity_loss}
\end{equation}
where $y_c$ is the ground truth classification label (see eq. (\ref{eqn:classification_loss})); $y_{h}$ is the ground truth $\log \mnhi$ label ($y_{h} = 0.0$ when no DLA exists, and in the range (19.5, 22.5) when a DLA exists); $\hat{y}_{h}$ is the model predicted $\log \mnhi$; and $\epsilon = 10^{-6}$ 
is a small value safe from 32-bit floating point rounding error.

The final loss function for our multi-task learning model is the sum 
of the 3 individual loss functions: 
\begin{equation}
\mathcal{L} = \mathcal{L}_{c} + \mathcal{L}_{o} + \mathcal{L}_{h}
\end{equation}
We note that classification loss is not actually used in determining the existence of a DLA. We get better results by analyzing the models ability to predict the location of the DLA as we make predictions across the sightline. Classification loss proved to be a good regularizer for the network, improving the results of both localization and log column density estimates; this is its sole purpose.

\subsection{Masking the gradient} 
\label{section:maskingthegradient}
There is a conceptual problem with the neural network described previously: 
on negative samples (those with no DLA in the 400 px window) 
the column density measurement has no meaning and the adopted label is 0.0. 
But if the model were trained to predict 0.0 when it 
classifies the regions with no DLA present, this would 
induce a bias towards 0.0 on positive samples because fully 50\% of the training data (negative samples) would be spent predicting 
a column density result of 0.0.

To get around this problem, we mask the gradient of column density during 
training when the sample is negative. 
This is achieved by adding a mask term to the 
loss function for column density. The original loss function 
for column density is the standard square loss; we 
multiply this by a constant term that is 0 for a negative sample 
(no DLA) and 1 for a positive sample, see equation (\ref{eqn:columndensity_loss}). 

We multiply in the term: $\left(\frac{y_c}{y_c + \epsilon}\right)$ where $\epsilon = 10^{-6}$, a value near zero that isn't affected by 32 bit floating point rounding error. $y_c$ is the classification label of the sample. The multiplicative term will be approximately 1 when the label is non-zero, and approximately 0 when the label is 0 (i.e. a negative sample). During back propagation, this constant term will carry through to the derivative, zeroing it out.

%
%
%

The three loss functions are summed and the ultimate effect is that the column density loss does not contribute to the gradient for regions without a classified DLA.
At inference time a column density value will be produced for negative samples 
by the model, but the value produced is simply ignored when no DLA is predicted 
and has no useful meaning.
The idea of masking gradients is common to recurrent neural network implementations, but not commonly applied to feed forward or CNNs.

\subsection{Overlapping DLAs} 
\label{sec:overlap}

Because the incidence of DLAs is low, most spectra contain 1 or 0 systems.
With a large enough dataset, however, sightlines with multiple DLAs
occur and a subset of these will contain pairs of DLAs close in absorption redshift.
We refer to these as ``overlapping'' DLAs.
Labeling samples where two DLAs exist near each other or overlapping involves changing the localization and column density values such that they refer to the nearest DLA.  This is a slight modification from what is seen in the visualization in Figure~\ref{labeling_diagram.pdf}. For example if the center of two DLAs is 30 pixels apart, labels for localization would begin at $-60$, and continue to the first DLAs center with a localization label of $0$, then begin counting up to $+30$, at which point it would flip to $-30$ and count down again to the second DLA's center with label $0$, and back up to $+60$. 
A similar approach is applied to column density. 
When calculating the column density measurement we normally average
column density measurements across the $\pm 30$ values around
the DLA center (see Section~\ref{-section:column_density_measurement}).
When two DLAs overlap we make use of fewer column density 
measurements such that we use column density measurements up to the center point between the overlapping DLAs.

\section{Training}
\label{sec:training}

\subsection{Considerations}

Figure~\ref{pos_neg_regions.pdf} shows a sample spectrum
and the positive (green), negative (red), and ignored (gray) regions. 
Notably there are far more negative regions than positive regions along 
the sightline. Training on more negative than positive samples would 
induce bias in the algorithm, so we balanced the positive and negative samples trained on.
By design, our training dataset (described in \ref{sec:training_set})
only contains sightlines with DLAs. 
In training we sample every possible positive samples from the sightline, 
and an equal number of negative samples are chosen from the sightline at 
random so that we maintain a 50/50 balance between training on positive 
and negative regions. This is standard practice in training such 
models\footnote{See \cite{Batista2004} for more information 
on balancing labels in machine learning.}.

Another issue arises at the boundary where 
a valid DLA is 61 pixels from the center of the window and the localization 
label abruptly changes from $\pm60$ to 0.  This would certainly cause 
trouble for the learning algorithm. 
Therefore, we do not train the model on these cases, and during inference 
we do not need the values at these edges to be accurate; they go unused and untrained.

We also ignore regions of the sightline where 
\Lyb\ absorption corresponding to identified DLAs takes place. 
The \Lyb\ absorption from DLAs may be falsely
detected as \Lya\ from gas at lower redshift.
The algorithm is not trained in a manner 
that would allow it to identify the difference between  
\Lya\ absorption and \Lyb\ absorption. Although this is 
potentially feasible, we did not include it in the scope of this work.
We simply compute the \Lyb\ location and mask any 
identified absorption as \Lyb\ in post-processing (any corresponding strong
HI absorber within 15\AA\ of the predicted location of \Lyb).
Training on these regions would not stop the algorithm from learning, 
but lowers its accuracy by training on labels that indicate no absorption exists despite
its presence. We do not train the algorithm on these regions, and remove them in post-processing at analysis time.
While this reduces the spectral range searched for DLAs, the effect is 
small and eliminates a significant source of false positives
from a DLA survey.

\img{}{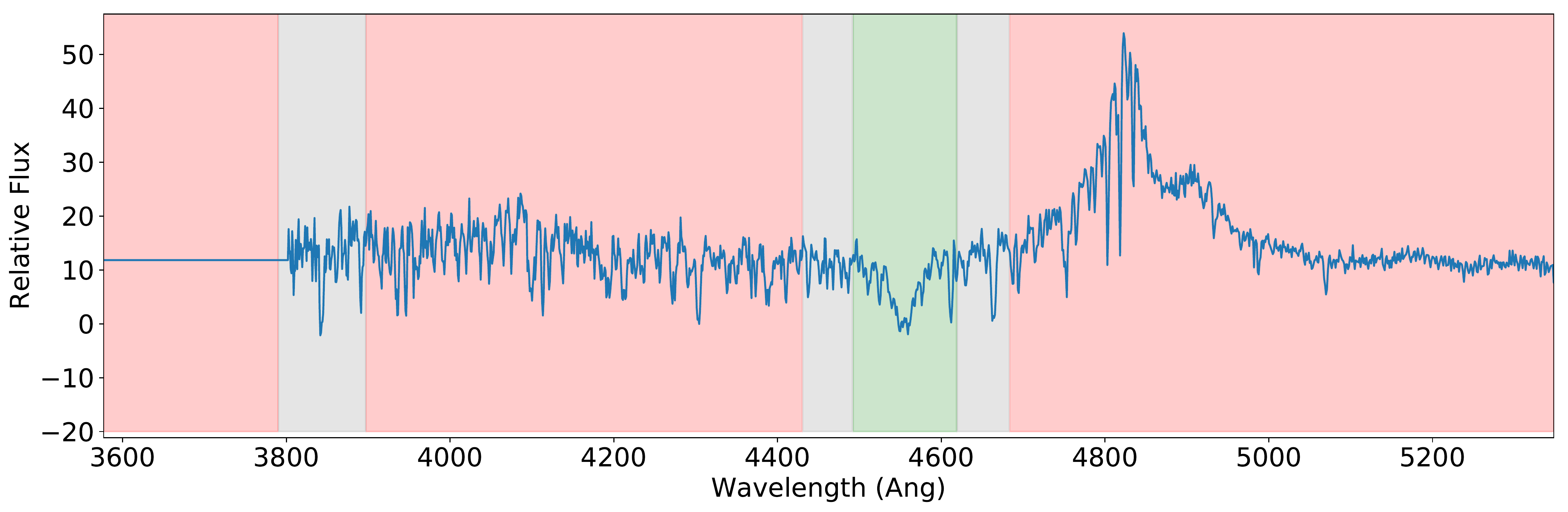}{A colored representation of positive (green), negative (red), and ignored (grey) regions of the sightline. The green regions indicate training samples that have a DLA, the red regions are training samples that do not have a DLA, and the grey regions are excluded. The grey regions near the positive DLA are cases on the border between a DLA existing in the 400\,pixel sliding window and not. The grey region at shorter wavelengths spans as the \Lyb\ absorption region of the DLA. For DLAs with high \nhi\ values, in particular, their \Lyb\ absorption mimics smaller DLAs and causes the training accuracy to decrease.}

\subsection{Training Set}
\label{sec:training_set}
In many respects, a CNN model is only as good as its training set
and this bore true for our DLA analysis.  Initially, we began with 
human-generated catalogs of DLAs \citep[e.g.][]{Noterdaeme2012} but 
found that these were too impure and too small 
($N \lesssim 10,000$ DLAs)
to train the CNN.  We moved, therefore, to generate our own
training set by inserting mock DLAs and SLLSs
into SDSS spectra that were
believed to have no DLAs.

Referring to the published survey of DLAs by \cite{Prochaska2009},
we identified the 4,113 sightlines that they
analyzed which had no reported DLA, had a quasar redshift
$\mzem > 2.3$, a S/N greater than 5 measured from the
median flux to error estimate near the quasar \Lya\ emission,
and no BAL signature (flag=0).
We will refer to these 4,113 sightlines as "DLA-free" and emphasize
that they can and do include SLLS.
We further note that this sample has higher S/N than
a random sampling of $z>2$ SDSS quasar sightlines, i.e.\ 
our model is not trained extensively in the low S/N regime.

In several stages, we proceeded to generate 200,000 sightlines
for training. For the first set of 135,000 sightlines, we drew
emission redshifts uniformly from the range
$z_{\rm em} = 2.3-4.5$ and then matched it to
the DLA-free sightline with smallest offset in \zem.  
We then inserted at least  one DLA\footnote{The injection
incidence was based on the observed incidence but we
drew repeatedly until at least one DLA was recovered.} 
into each sightline with column density 
drawn randomly from
the \nhi\ frequency distribution of \cite{Prochaska2005}
over the interval $\log \mnhi = 20.3-22.5$.
The absorption redshifts \zabs\
are constrained to place \Lya\ between
910\AA\ and 1215\AA\ of the quasar rest-frame.
A Voigt profile was generated for each absorber and multiplied
into the spectrum. This depresses the flux and also the noise
fluctuations in the original data.  
Therefore, we added random fluctuations (with normal deviate) 
within the Voigt profile based on the error array of the spectrum
and scaled by $1-f_V$ where $f_V$ is the normalized Voigt profile model.
An example of mock-generated DLAs is shown 
in Figure \ref{fig_dla_injection.pdf}.

To enable training on SLLS, we generated 45,000 sightlines
as above but with systems having 
$\log \mnhi = 19.5-20.3$.  
Lastly, we generated 20,000 sightlines with DLAs 
having $\log \mnhi = 21.2-22.5$.  This high \nhi\ sample 
improved the model's ability to characterize
the strongest DLAs.
We note that these training spectra ignore several aspects
of real data: 
(i) metal-line absorption;  the
majority of DLAs show strong, associated metal-line
transitions including saturated SiIII~1206 absorption
near \Lya;
(ii) any clustering of \Lya\ lines to DLAs as may be
expected from hierarchical cosmology.  We believe, however,
that ignoring these effects has a minor impact 
on our results.

Training iterates through 100s of millions of data points, 
each being a random 400-pixel segment drawn from the above sightlines, 
also randomly chosen with the following exceptions:  
(i) the approximately 2,000
sightlines that have overlapping DLAs were sampled at 
$4\times$ the otherwise random rate to improve the algorithm's
ability to resolve these difficult configurations;
(ii) the SLLS sightlines
were sampled at half the rate as those with DLAs to limit
their impact.; and
(iii) positive/negative segments are selected at a rate of 50/50, 
i.e. we sample more densely from the positive 
regions of the sightline which 
are smaller segments than the negative regions (Figure~\ref{pos_neg_regions.pdf}).




\imgsinglecol{}{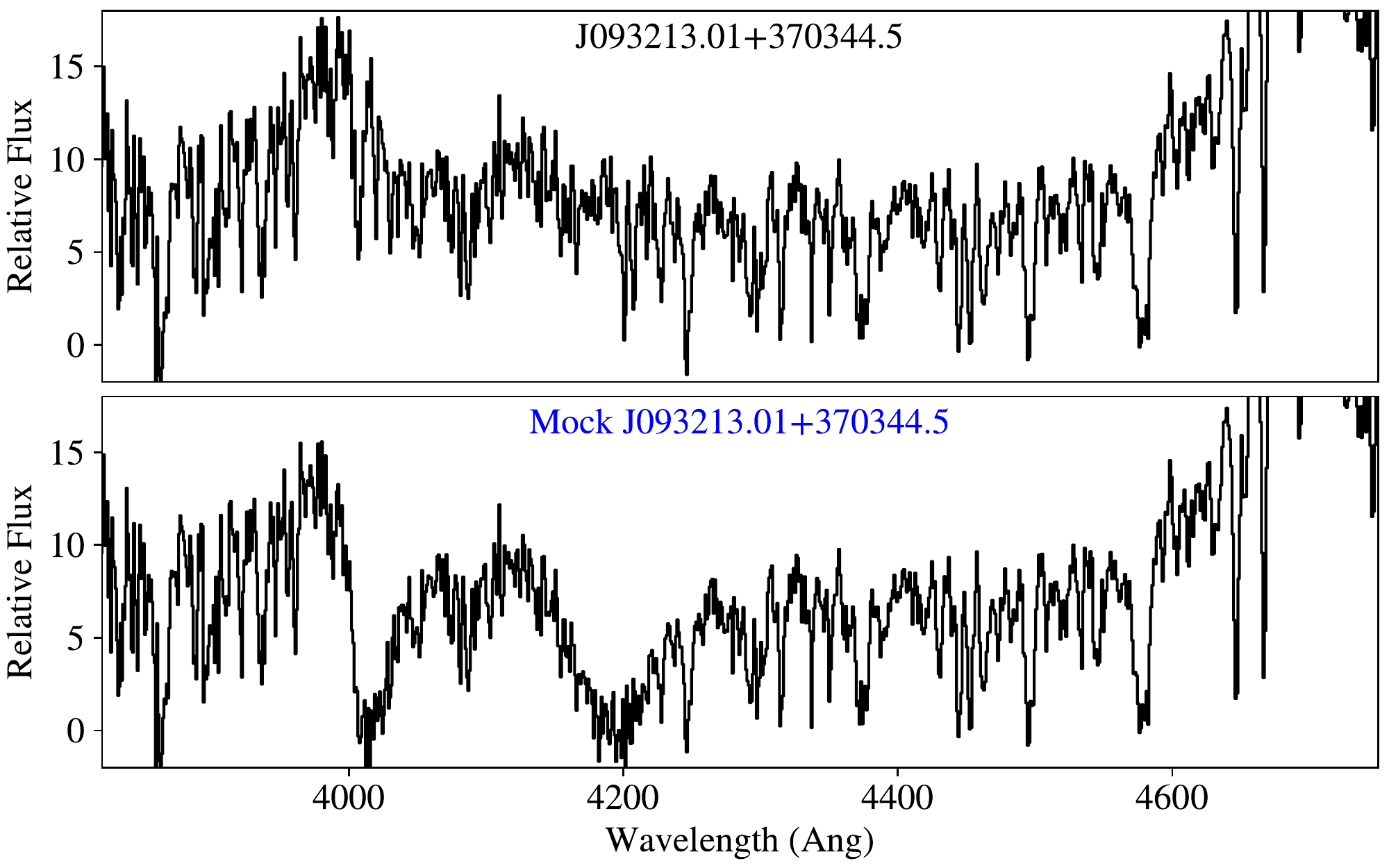}{(top) The SDSS DR7 spectrum
of J093213.01+370344.5 which was characterized as DLA-free 
by Prochaska \& Wolfe (2009); 
(bottom) the same sightline with two injected DLAs at 
$\mzabs \approx 2.303, 2.450$ and with $\log \mnhi \approx 20.5, 21.0$ respectively.}



Lastly, 10,000 sightlines with at least one injected DLA (and no 
injected SLLS) were held for validation (see Section~\ref{section:validation}).

\subsection{Hyperparameter selection} 
\label{section:hyperparameter_selection}
Hyperparameters were selected by identifying those
that produce the best results on a held out validation dataset. 
There are 23 hyperparameters -- everything from L2 regularization penalty to kernel sizes. 
All hyperparameters are listed in table~\ref{-table:hyperparameters}. 
Hyperparameter selection was performed using a restricted coordinate descent method. Initially a reasonable value for the hyperparameters was configured manually. Then a number of values above and below this value were selected.
We then randomly selected one hyperparameter and 
trained the network using each possible value for the hyperparameter. 
The value that yields the model with the best validation error is kept as 
the parameter value. 
We then choose a new hyperparameter and repeat the process until reasonable convergence (usually a day or two worth of training on our 3 layer convolutional network described in section \ref{section:convolutional_network_architecture}). For hyperparameter search we ran training using two NVIDIA K20 GPUs to try different parameter combinations and limited the training to approximately 30 minutes, which was a reasonable trade-off between the final 1-2 day convergence and training time.

Through this process we were surprised by a few hyperparameter values. 
First, the number of filters selected for each layer of the network was very
large.  More than 100 filters in the first convolutional layer, for example, produced over-fitting. The batch size was a hyperparameter that was selected and the chosen value was clearly affected by the amount of data analyzed. 
When there was a large amount of data and over-fitting was less of an issue, a larger batch size produced better results. When we used smaller datasets a smaller batch size was used. 
The dropout keep probability and L2 regularization penalty behaved in a similar manner to batch size. Dropout was rendered all but ineffective as we grew the amount of data in our training set. The optimal kernel sizes of 32 and 16 were much larger than we had anticipated based on experience with image datasets where 3, 5, and 7 kernel sizes are common. The number of neurons in each fully connected layer was also subject to over-fitting when larger values were tested.

An unlisted hyperparameter that was originally tested is whether max pooling or mean pooling should be used. We never found a case where mean pooling outperformed max pooling and we eventually removed it as a hyperparameter.

\section{Validation}  
\label{section:validation}
To quantify the algorithm's ability to properly identify and measure 
DLAs we compare it to two catalogs: 
one where we inject DLAs into
real spectra believed to be DLA-free, and 
one manually analyzed by human experts on real data (\cite{Prochaska2009}). 
Both of these datasets have the same selection criteria for
quasar sightlines as the training set except being DLA-free.
We analyze the accuracy in terms of 
(i) the detection rate of DLAs; 
(ii) the measurement precisions of \zabs\ and \nhi;
(iii) the rate of false negatives, 
DLAs known to exist a priori that were missed by the algorithm;
and
(iv) the false positive rate, 
DLAs that were identified by the algorithm but were not previously
known. 

\subsection{DLA Analysis}
\label{sec:dla_analysis}

We begin by describing the application of the CNN model
to a single sightline for the identification and
characterization of DLAs.


\subsubsection{Preprocessing}
\label{-section:preprocessing}
The only preprocessing step on the spectral data is to 
pad the flux array on the left 
to create a fixed length sightline. This fills out each 
array to the same dimension but has no 
consequence to the algorithm. 
Padded values take on the mean of the first 50 pixels in the sightline, 
an arbitrary but effective choice.

\subsubsection{Processing a Spectrum}
\label{-section:processing_sightline}
The spectrum fluxes are replicated into 1,748 segments of 400-pixels each. 
The neural network receives each 400 pixel segment and, for each segment, 
outputs three results: 
1) A predicted classification in the range $(0,1)$\footnote{The output of the network is in the range $(0,1)$ whereas the known ground truth label is $\in \{0,1\}$} 
indicating whether a DLA exists within $\pm 60$~pixels 
from the center of the window; 
2) an offset value from $[-60, +60]$ which indicates the relative 
location in pixels of the DLA  to the center of the window 
(and 0 if the DLA is not within $\pm 60$~pixels of the center of the window); 
and 3) a logarithmic column density measurement ($\log \mnhi$). 
An example of the labels for a single sightline are
shown in Figure~\ref{fig_labels.pdf}.
On a multi-core Intel based architecture, the algorithm process 
500 spectra in approximately 140\,s using 19 cores in parallel.

\imgsinglecol{}{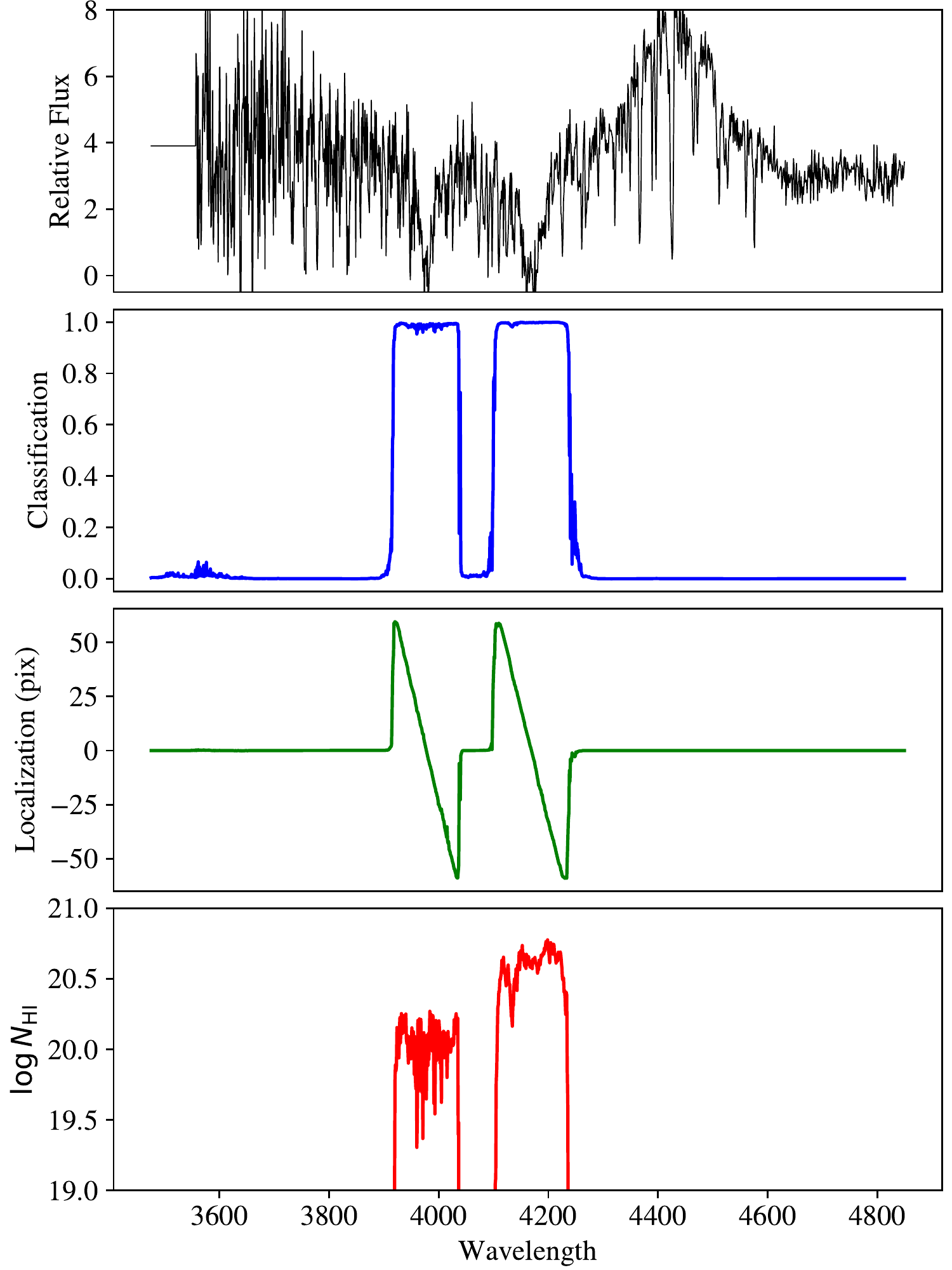}{Example showing the labels
along a single sightline.  From top to bottom, the panels show: 
(i) the relative flux values for the spectrum;
(ii) the classification labels;
(iii) the localization labels;
(iv) the \nhi\ labels.  Note that
the \nhi\ labels have no meaning (i.e.\ may be ignored)
for spectral 
regions outside where a DLA has been classified.
}

\subsubsection{Localization and Confidence}
\label{-section:localization}
Given the model results for the full sightline, 
the next task is to identify the center location of any
DLAs,  i.e.\ the absorption redshift $z_{\rm abs}$. 
In the example given by figure~\ref{fig_labels.pdf}, 
two DLAs were identified at $\approx 3980$\AA\ and $\approx 4170$\AA\
corresponding to $\mzabs \approx 2.27$ and 2.43. 
The full scan provides a set of offset values
which we histogram, expecting to observe
a cluster of values at the center of a bonafide DLA. 
These are shown in the lower panels of
Figure~\ref{fig_dla_confidence.pdf} 
as green points, where the value
indicates the fraction of the nearest 121 pixels ($\pm 60$)
that indicate the DLA is located at this wavelength, scaled by 1.5
to bring typical values of true DLAs close to 1.
We further define a \confp\ parameter
where we sum the histogram over the nearest 5 pixels 
(i.e.\ $\pm 2$ pixels)
and take a 9-pixel median filter of that result
limiting its maximum value to 1. The choice of a 9-pixel kernel 
for the median filter is arrived at by balancing between 
over and under smoothing and determined by experimentation
with the training set.

For well-localized DLAs, this \confp\ parameter shows
a sharp peak, as seen in figure \ref{fig_dla_confidence.pdf}. 
The spectrum is then scanned for peaks with \confp\ exceeding a 
threshold of 0.2. 
Any such peak signifies a highly probable DLA candidate. 
We emphasize that the main results of this manuscript are
insensitive to the choice for any threshold value within
0.2 and 0.7.
We mask all pixels with \confp\ $> 0.1$ that
neighbor the DLA candidate, and 
then search for additional DLAs in the spectrum. 

\imgsinglecol{}{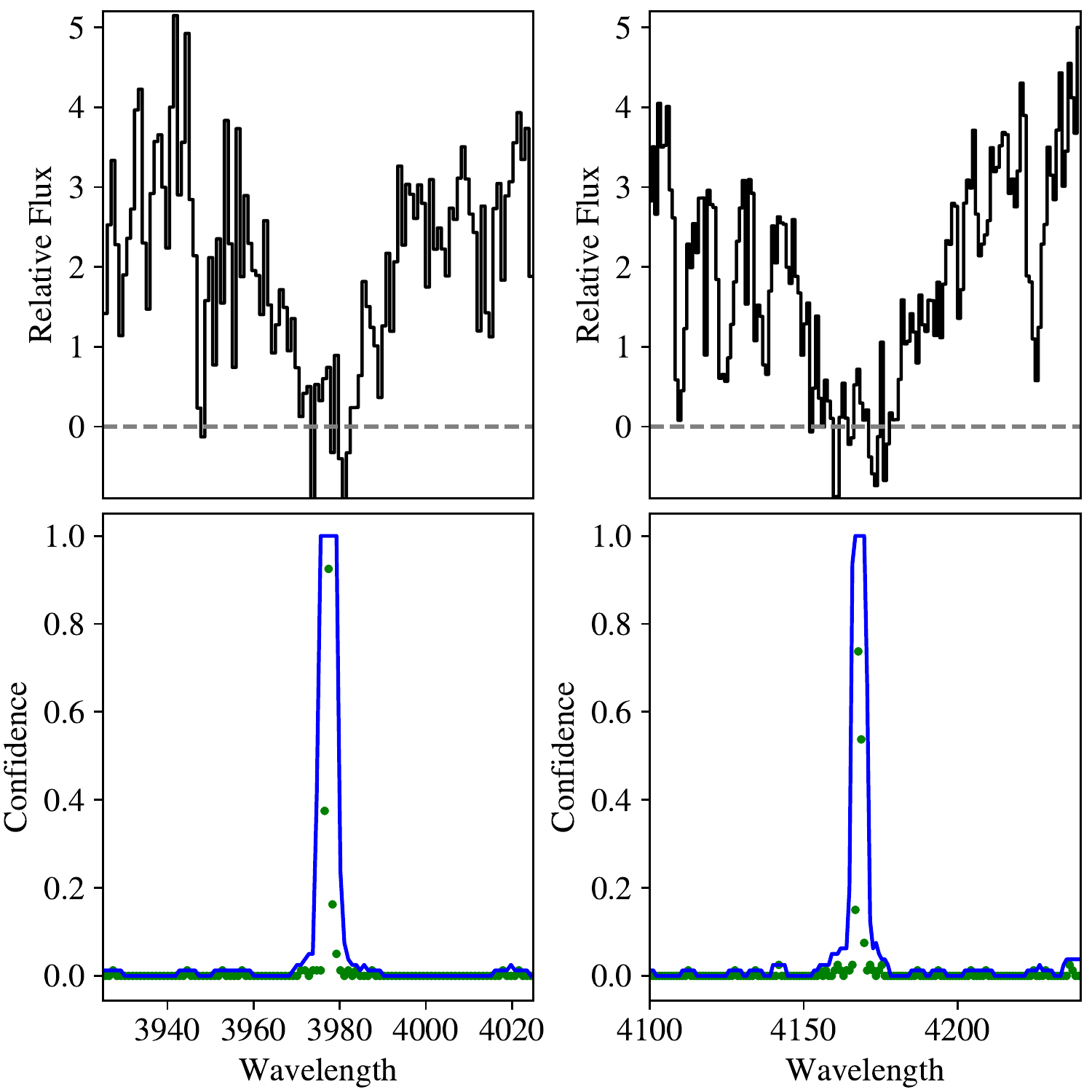}{
(top panels) Zoom-in snapshots of two DLAs classified in
the spectrum shown in Figure~\ref{fig_labels.pdf}.
(lower panels) Colored dots show at each pixel
the histogram of offset values within 
$\pm 60$\,pixels that predict the DLA
is located at the given pixel
(scaled by an arbitrary 1.5 factor).
The blue curve is the \confp\ parameter,
the 9~pixel median of the histogram that
is first summed in a boxcar of 5 pixels.
}

\subsubsection{Column Density Measurement}
\label{-section:column_density_measurement}

As with localization, we analyze multiple predictions for the
DLA \nhi\ value to derive the adopted value.
When a DLA is identified, we take the 40 measurements of $\log \mnhi$ 
to the left and right of the DLA and average them. In this case the choice of 
40 was a reasonable, but arbitrary number smaller than the $\pm$ 60 pixels used in localization/classification. 
The boundary at $\pm$ 60 pixels, where the label abruptly changes to 0,
is expected to be problematic for training and we avoided that region.
The bottom plot in Figure~\ref{fig_labels.pdf} shows
log column density measurements around each DLA. 
One notes a modest variance to the measurements. 
By analyzing multiple measurements, this 
(1) reduces scatter in the algorithm's 
prediction of log column density; 
and (2) the measured standard
deviation gives a crude indication of the 
uncertainty in the \nhi\ value. 
%


In all previous human-led works, column density measurements of
DLAs were derived from fits of a Voigt profile.  Our approach with deep
learning, however, does not explicitly 
implement any concept of Quantum Mechanics yet yields excellent results.
We suspect, however, that the algorithm learns by matching patterns such 
as the shape of the wings inherent in a Voigt profile to map to 
real-valued column density output.

The algorithm does exhibit a small consistent bias at the high and low 
end of \nhi\ values which we characterized using
a 5,000 sample validation dataset. 
A plot of the column density predictions and actual values on this 
validation dataset can be seen in Figure \ref{fig_test_nhi.pdf}. 
This distribution of bias is consistent across both generated and 
human tagged datasets. For the final measurements, and those reported here, 
we fit the bias with a simple 3rd degree polynomial 
ridge regression model, and add the bias from all samples. 
The 4 ridge regression coefficients are:
$f(x) = 0.0028149x^3 - 0.06461880x^2 - 0.004256562x + 23.5553179$, 
where $f$ is the amount of bias to 
remove from the models estimate, 
and $x$ is the biased column density estimate from the model. 

\imgsinglecol{}{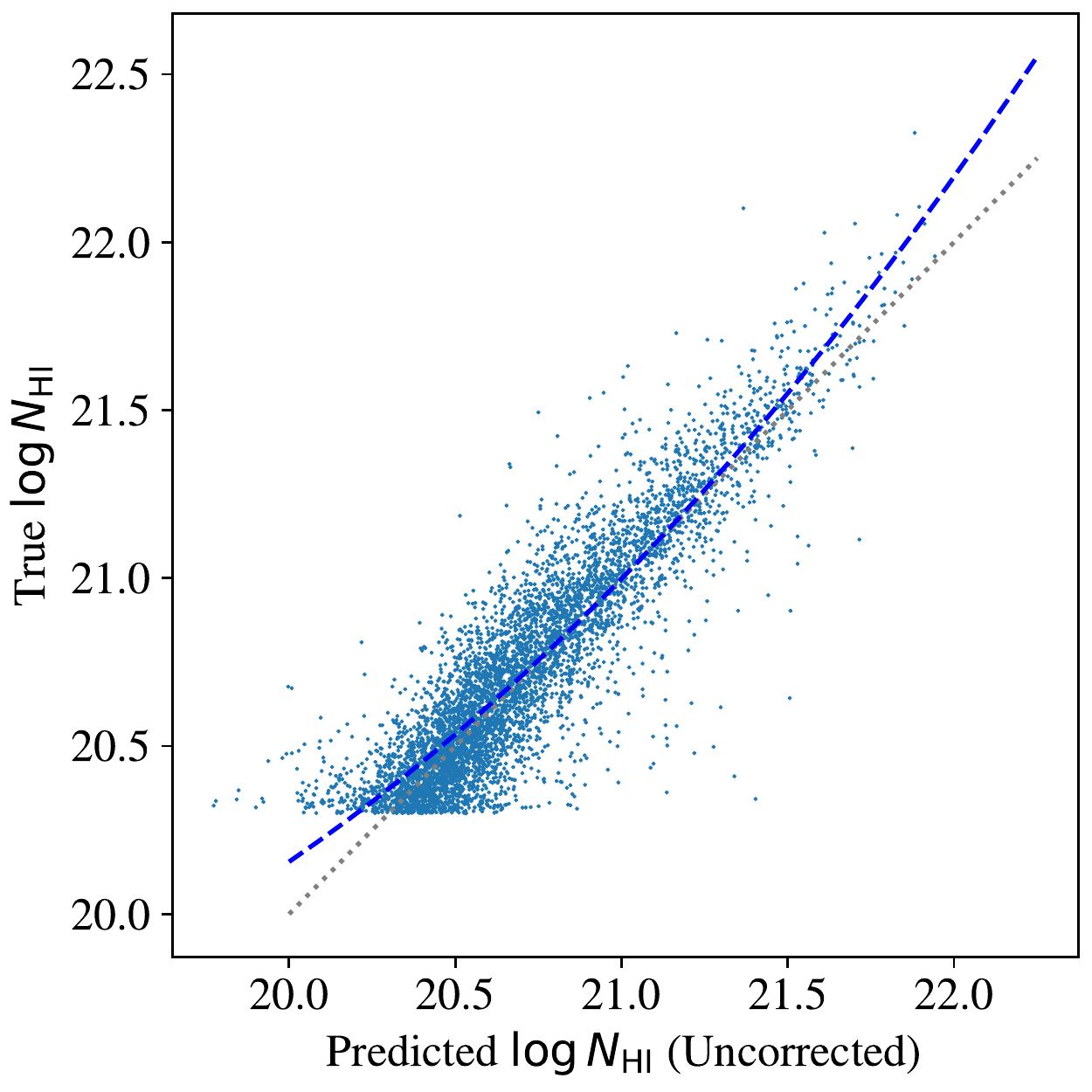}{Scatter plot of 
actual column densities of injected DLAs versus
the uncorrected \nhi\ value predicted by the algorithm.
The gray dotted curve is the one-to-one line marking a
perfect prediction. 
The blue dashed-line is a 3rd degree polynomial ridge regression fit to the 
points, showing a small negative bias, more pronounced at the large 
and small extremes.  This fit is applied to all predicted \nhi\ values.}

\subsection{Comparing results against a generated catalog}
\label{-section:results_comparison_gensample}

The results listed in this section report on the algorithm's ability to identify and 
measure DLAs for which we know \zabs\ and \nhi\ exactly.  
Specifically, we analyzed 10,000 sightlines
from the training set that were held out.
Each sightline in this dataset contains 1 or more artificially injected 
DLAs ($\sim$87\% have 1 DLA, $\sim$12\% have 2 DLAs, $\sim$1\% have 3 DLAs, 
and only a few have 4 DLAs), for a total of 11,403 generated DLAs.

\subsubsection{Matched DLAs}
To match our algorithm-predicted DLA with those injected, 
we require the predicted absorption redshift lie within 
$\delta z = 0.015$ ($\approx 1200 \, \rm km \, s^{-1}$)
of the true value.
With this $\delta z$ criterion, the algorithm successfully recovered 
11,110 DLAs from the test set, i.e. 97.4\%.  Furthermore, an additional
184 of the true DLAs were classified by the algorithm
as an SLLS (i.e.\ a system having $\log \mnhi < 20.3$).
This is expected and gives an overall strong HI absorber yield of 99\%.

For the 11,110 matched DLAs, we measure a median $\delta z$ of $-0.00037$
($\approx -30$ km/s) and a standard deviation of $\sigma(\delta z) = 0.0022$.
The distribution of log column density difference 
$\mdnhi \equiv \log \mnhi^{\rm true} - \log \mnhi^{\rm model}$
forms a Gaussian distribution with mean 0.031 and standard deviation 
of 0.15\,dex.

%
%
%
%

\subsubsection{False Negatives}
\label{--section:false_negatives}

In the 10,000 sightlines tested, a total of 299 injected DLAs were
not identified as a DLA by the algorithm.  
Figure~\ref{fig_test_false_neg.pdf} shows a scatter plot of 
\zabs\ and \nhi\ for these false negatives on a 
2D histogram of the entire injected distribution.  
Approximately 2/3 of the false negatives
were classified by the algorithm
as an SLLS (i.e.\ with $\log \mnhi < 20.3$) 
almost all have a truly low \nhi.  
This behaviour is both expected and reasonable.  

\imgsinglecol{}{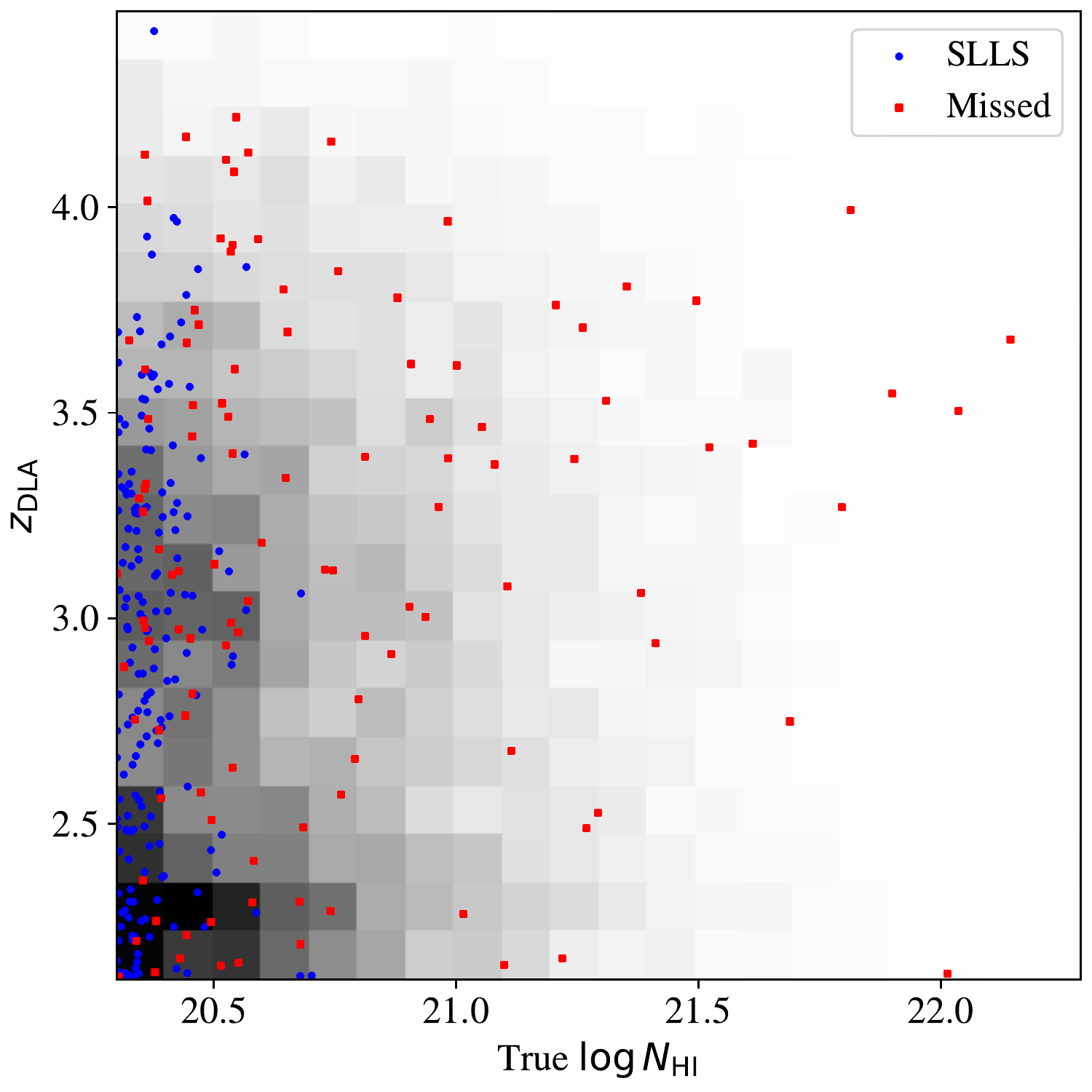}{Scatter plot of \zabs\ versus \nhi\ 
for the false negatives from the 10,0000 sightline validation
set.  The underlying gray histogram illustates the distribution of \zabs,\nhi\
pairs for the full, injected sample of 11,143 DLAs.  
Approximately 2/3 of the false negatives are characterized as
SLLS by the algorithm and have true $\log \mnhi < 20.5$\,dex.
The false negatives that were `missed' by the algorithm
(red squares) includes a set of $\approx 20$ systems with large,
true \nhi\ values.  See the text for a discussion of these which
are dominated by overlapping DLAs.
}



The remaining 115 cases of false negatives that were missed
by the algorithm occur for a number of reasons.
The dominant cause is overlapping DLAs, i.e. two or more 
DLAs with nearly coincident redshift. 
As is apparent from Figure~\ref{fig_test_false_neg.pdf},
the missed DLAs are biased towards high \nhi\ value.
This results from the fact that the algorithm 
occasionally mis-classifies 
(i) an injected DLA with high \nhi\ 
as two or more DLAs;
(ii) a pair of injected DLAs as a single DLA;
and
(iii) a pair of injected DLAs as two offset DLAs. 
Examples of the first two cases
are shown in Figure~\ref{fig_test_neg_overlap.pdf}.

While the algorithm properly predicts
overlapping DLAs in the majority of cases,  
there are still a non-negligible
incidence of mis-classifications.  Of course, 
human-based approaches are also prone to this issue. 
These high \nhi\ false negatives constitute 20 of the
1200 injected DLAs with high value ($\approx 2\%$).
We consider this an acceptable rate but will consider
a larger training set of DLAs with high \nhi\ to 
(hopefully) improve learning on these systems.

\imgsinglecol{}{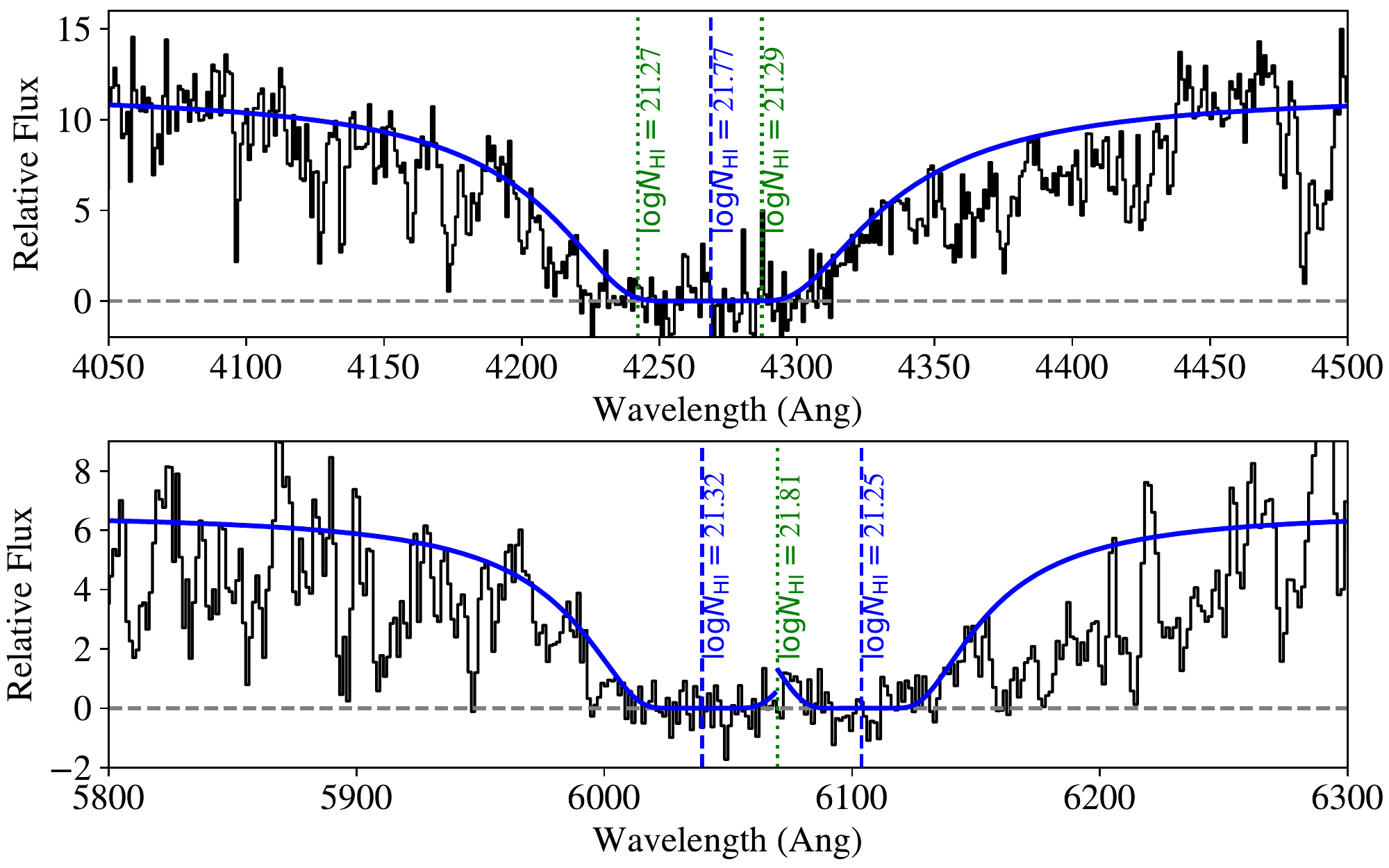}{
(top) An example of two overlapping DLAs (marked by green dotted
lines) generated as part of the training and test datasets, which the algorithm erroneously identified as one DLA (blue, dashed line). 
Note that the total \nhi\ value predicted is comparable to the
sum for the pair of DLAs.
(bottom)
An example where the algorithm predicted that a single, injected DLA
is a pair of overlapping DLAs.  Both of these cases lead to false
positives and negatives in the analysis.} 


The next most frequent cause of missed false negatives 
was for DLAs injected within the noisy 
end of the spectrum, i.e.\ typically below $\approx 4000$\A. 
The failures here have a lower signal to noise ratio, 
and in many cases even a human expert would struggle to identify the DLA. 
By the same token, we identify examples
where the algorithm successfully identifies a DLA in 
S/N$ < 2$\,per pixel conditions (Figure~\ref{fig_test_low_s2n.pdf}). 
Our assessment is that
the algorithm performs at least as well 
as a human expert in these low S/N cases. 

Completing the census of false negatives, 
there were 8 cases in which an injected DLA 
overlapped a \Lyb\ absorption and was ignored.
Such cases will exist in real datasets and are an unavoidable issue.
In six other cases the algorithm did in fact identify the DLA, but failed to measure 
its center point to within $\delta z = 0.015$. 
Finally there are three cases where the algorithm 
ignored the injected DLA for non-obvious reasons.

\imgsinglecol{}{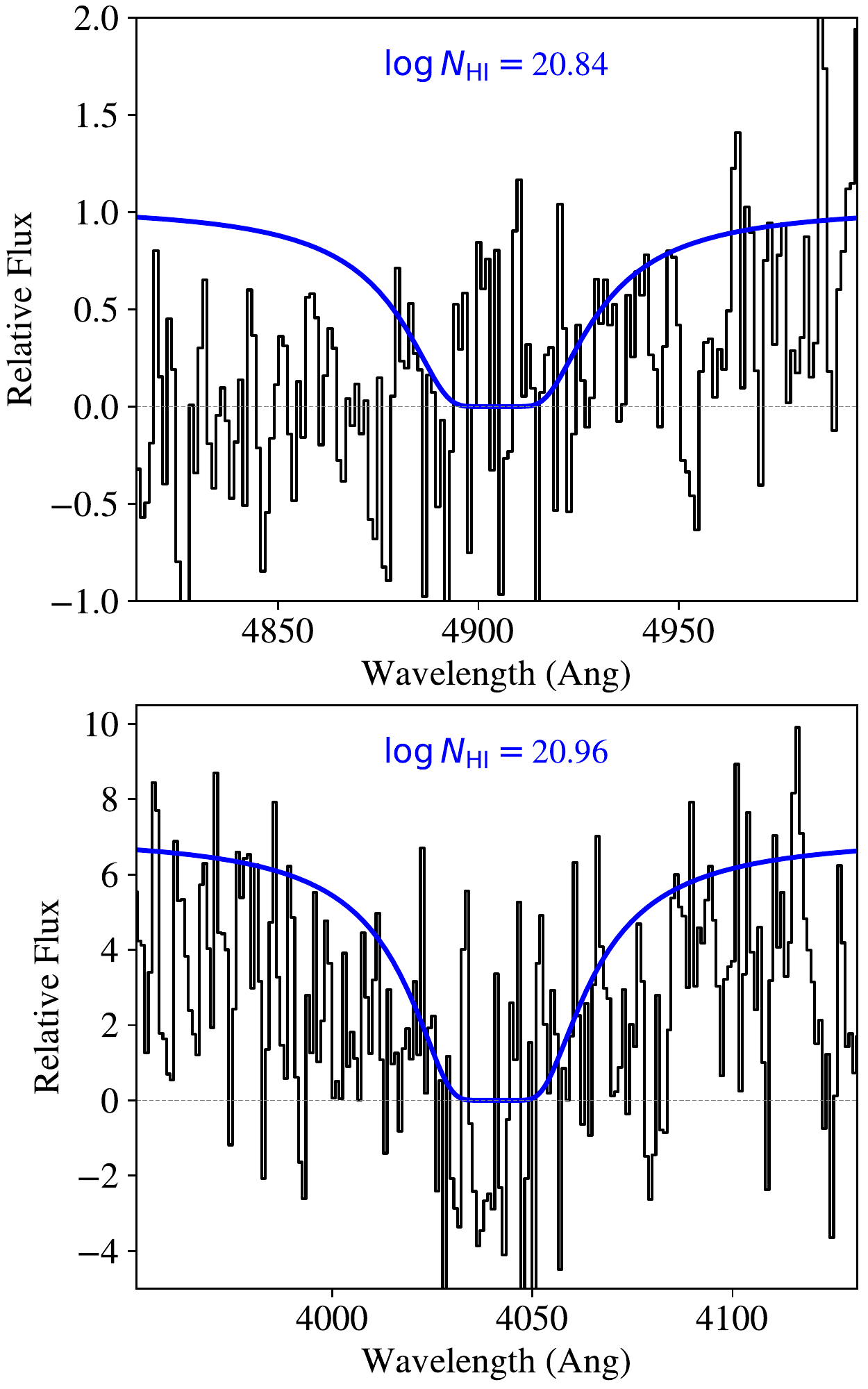}{
Two examples where a DLA was manually injected into a low S/N region
of spectra and correctly identified by the algorithm.  Very few humans
would confidently identify either of these systems.}

\subsubsection{False Positives}
\label{--section:false_positives}

Generally, a false positive is defined as a reported DLA
that does not match any injected DLA.  
By that definition, the algorithm identified 158 DLAs
without a match within $\delta z = 0.015$ to the injected
sample.  However, the DLA-free SDSS sightlines used to build
the training set may include DLAs in 
the spectral region not surveyed by \cite{Prochaska2009}.  These are the
bluest, lowest S/N regions of the spectra.  Indeed, 83
of these putative false positives fall within these 
non-surveyed spectral regions and a visual inspection
reveals that many ($>40$) 
are likely  legitimate DLAs.  Therefore, we restrict our definition
of false positives in this validation as those within
the statistical \cite{Prochaska2009} survey.


With this definition, 74 false positives were identified in 
this dataset of 10,000 test sightlines with 11,143 injected DLAs. 
Of these, only 1/3 have a high \confp\ value ($>0.9$).
We have manually inspected these 74 false positives and have categorized them as follows:
%
%
31 cases have low \nhi\ values (i.e.\ near the DLA threshold) and
it is possible that a fraction of these are legitimate DLAs. 
Indeed, many of these are listed within the DR5 catalog 
of \cite{Prochaska2009} as SLLSs.
A few typical examples of these false positives are
shown in figure \ref{fig_test_false_pos.pdf}. 

\imgsinglecol{}{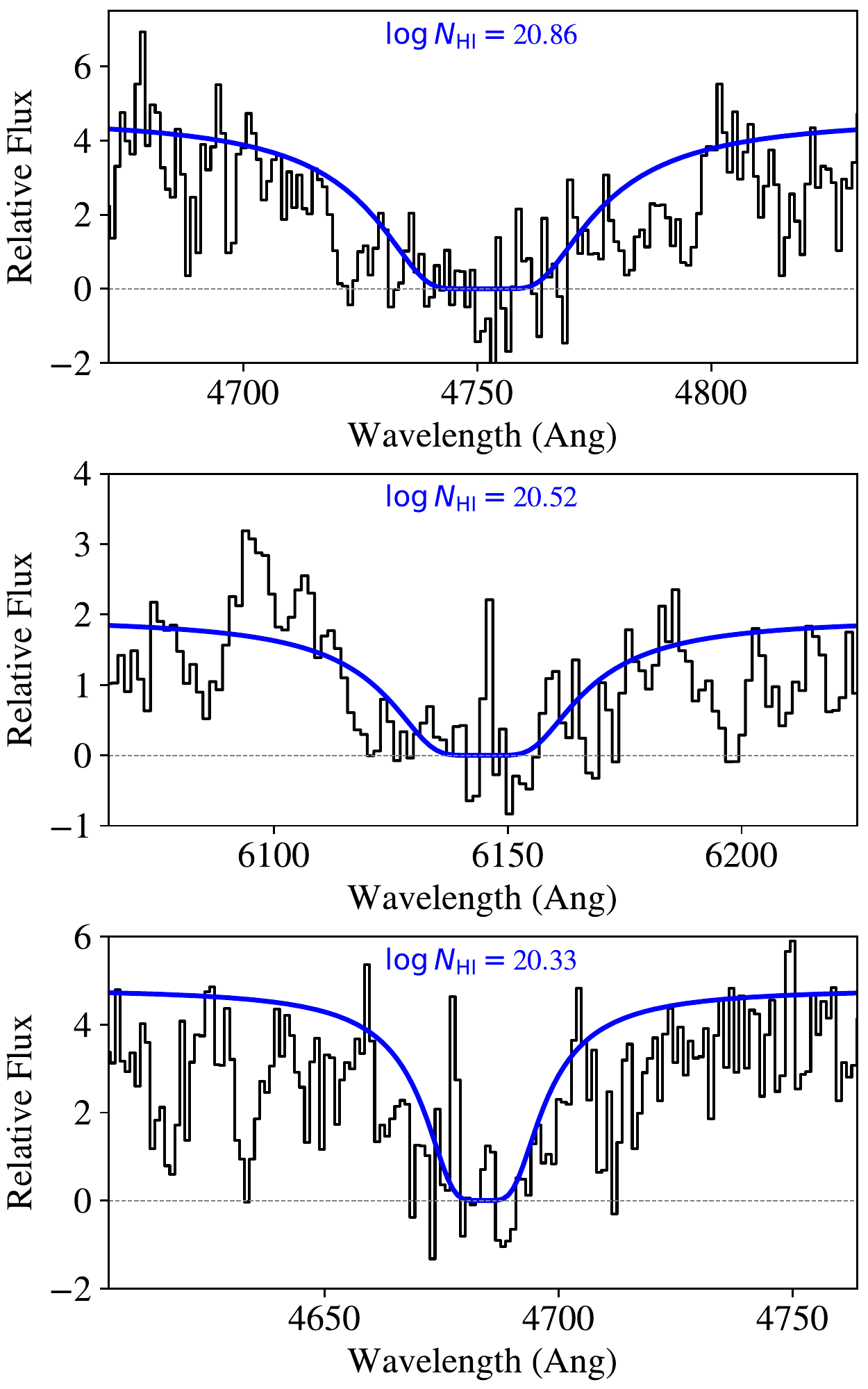}{Examples of typical false positives 
in the generated validation set.  The majority of appear to be
bonafide DLAs in the purported DLA-free sightlines of 
Prochaska \& Wolfe (2009); 
very few are egregious errors of the algorithm.}

Another 25 of the false positives are due to errors related 
to overlapping DLAs with one or both having high \nhi. 
Approximately half are cases where the algorithm identifies a 
single injected DLA as two absorbers and the other
half are two injected DLAs reported as single one with a 
significantly offset redshift (Figure~\ref{fig_test_neg_overlap.pdf}). 
The remainder result from a miscellaneous range of errors
(e.g.\ a strong \Lyb\ line misidentified as a DLA)
with only a few egregious cases.
The false positive rate reported here is very low, but 
we caution that this validation set did not include injected
SLLS which would increase the rate.



\subsection{Generalizing to real-world data: DR5}
\label{-section:generalizing_to_DR5}
Our algorithm performed well on the 10,000 sample test set, 
drawn from the distribution of sightlines used for training.
This might, however, be considered too idealized of
a test for validation. 
To further test our algorithm, we examined how the 
same model generalizes to real-world sightlines. 
Specifically, we analyzed the same SDSS sightlines surveyed
by \cite{Prochaska2009} and restricted to their statistical sample of DLA.
We further limited our analysis to the SDSS DR7 
quasar spectra catalog provided in the {\it igmspec} database
\citep{igmspec}
which eliminates five sightlines from \cite{Prochaska2009}.   Altogether, 
this validation set includes
737 DLAs in 7477 sightlines.

\subsubsection{Results for Matches}
\label{sec:dr5_matches}

Our algorithm predicted a DLA matched to within $\delta z = 0.015$
for 670 of the 737 statistical DLAs of \cite{Prochaska2009}.  Figure~\ref{fig_dr5_vs_ml.pdf}
shows histograms of the offsets in redshift and \nhi\ between the 
algorithm's predictions and the \cite{Prochaska2009} values.
The median redshift offset is only 0.0007 with a standard
deviation of 0.0028.  The
median log column density offset is $\Delta \mnhi  = -0.006$
with standard deviation of 0.18\,dex, 
quite similar to the generated training and test datasets. 

\imgsinglecol{}{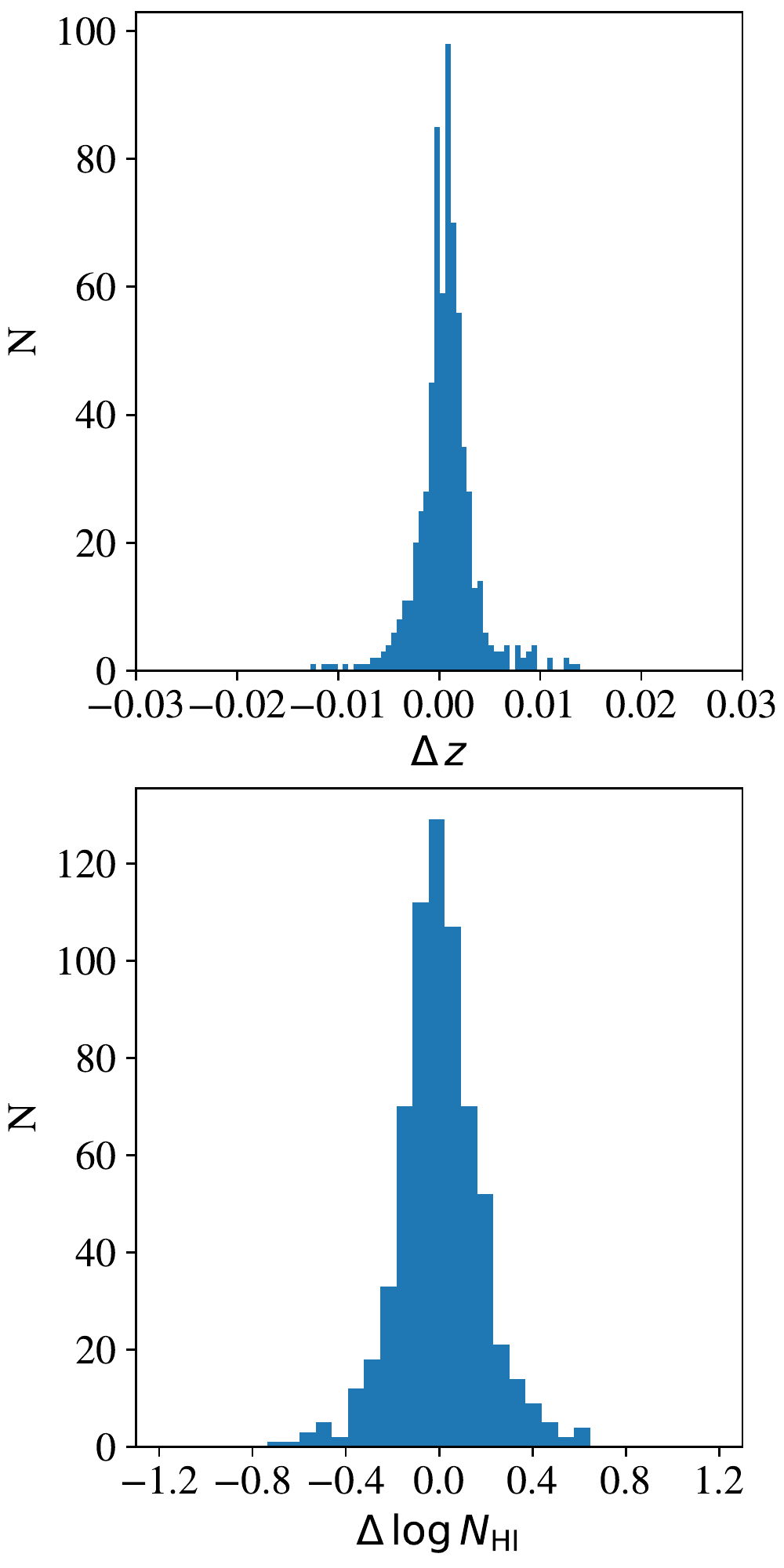}{
Redshift (top) and \nhi\ (bottom) offsets
for the DLAs matched between the algorithm
and the human-identified DLAs in the SDSS DR5
\citep{Prochaska2009}.  The rate of DLA recovery by the
algorithm is very high ($>95\%$) and the
accuracy in \zabs\ and \nhi\ is excellent.
}


%
%
%
%
%
%
%

\subsubsection{False Negatives}
\label{--section:dr5_false_negatives}

False negatives are defined as DLAs in the statistical DLA sample of 
\cite{Prochaska2009}
that were not identified by the algorithm. 
There are 67 such cases and 49 of these were 
reported by the algorithm to be an SLLS.
These false negatives are expected and the rate is consistent
with expectation.
The remaining 18 are a combination of redshift offset
$\delta z > 0.015$, 
overlapping DLAs, and several egregious misidentifications.
Allowing for the 49~SLLSs, the algorithm discovered a total
of $719/737 \approx 97.6\%$ strong HI absorbers
from the set of human-identified DLAs.

\subsubsection{False Positives}
\label{--false_positives}
False positives account for cases when the algorithm identifies a DLA 
that was not previously labeled in the DR5 dataset. 
As with the test set, we restrict the definition to DLAs in the
spectral regions surveyed for DLAs by \cite{Prochaska2009}. i.e.\ 
regions that satisfied their S/N threshold\footnote{We note that the
algorithm recovered 811 DLAs outside this region that were
not reported by \cite{Prochaska2009}.}.
There are 191 false positives that fell 
within the range analyzed by \cite{Prochaska2009}.
Of these, 111 have $\log \mnhi\ \le 20.45$, i.e. near the DLA
threshold which may well be true DLAs.   
Because these false positives outnumber the false negatives, this
could indicate a several percent underestimate in the
incidence of DLAs by \cite{Prochaska2009}.

There were six false positives 
with $\log \mnhi > 21$ which we examined closely.  
Three of these are obvious DLAs
that were not reported by \cite{Prochaska2009} for unknown reasons ($!$)
and two are proximate DLAs.
The sixth occurs at $z \approx 5$ in a noisy spectrum 
and is likely erroneous.
There are 35 false positives with intermediate \nhi\ values
($20.6 < \log \mnhi < 21$).  Of these, 27 occur along sightlines
with $\mzem > 3.8$ and lower S/N spectra.  By human inspection,
most are plausible candidates but we suspect the algorithm is 
over-predicting the incidence.  The other eight are high quality
systems that were missed by \cite{Prochaska2009}.





We highlight a particularly interesting false positive
case where the algorithm identified a DLA 
that at first glance looks to be in error. In figure \ref{noise_in_dla_case.pdf} a DLA is identified in the DR5 dataset 
despite a notably significant positive flux in 
the middle of the DLA. The SDSS DR5 catalog 
did not indicate any error with this flux, and thus we first considered
the DLA to be a false positive.
However analyzing this same sightline in the BOSS-DR12 dataset 
reveals that the flux at $\lambda \approx 4778$\A in the SDSS
spectrum was spurious and that a true DLA is present!  
This is notable because these cases are learned, the algorithm identified a DLA in this case because there are examples of night sky flux in 
the training samples. 


\imgsinglecol{}{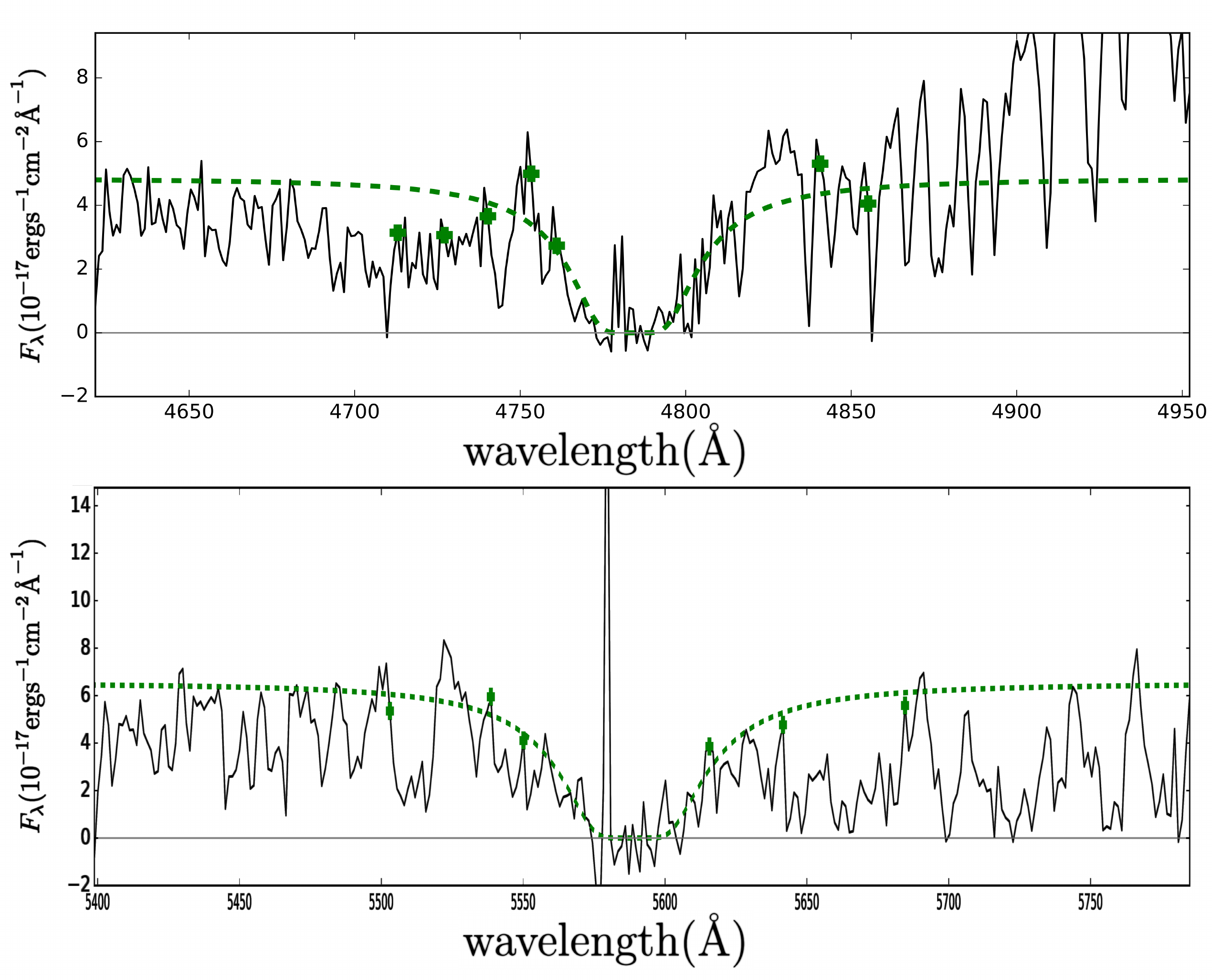}{\textit{Above:} 
Sightline from DR5 (plate 2111, fiber 525) with a reportedly relevant flux in the middle of the DLA which is identified as a DLA only
by the algorithm. \textit{Below:} An example from training data with 
night sky flux coincidentally part of the DLA.  This absorption
was also properly identified as a DLA by the algorithm.}

\imgsinglecol{}{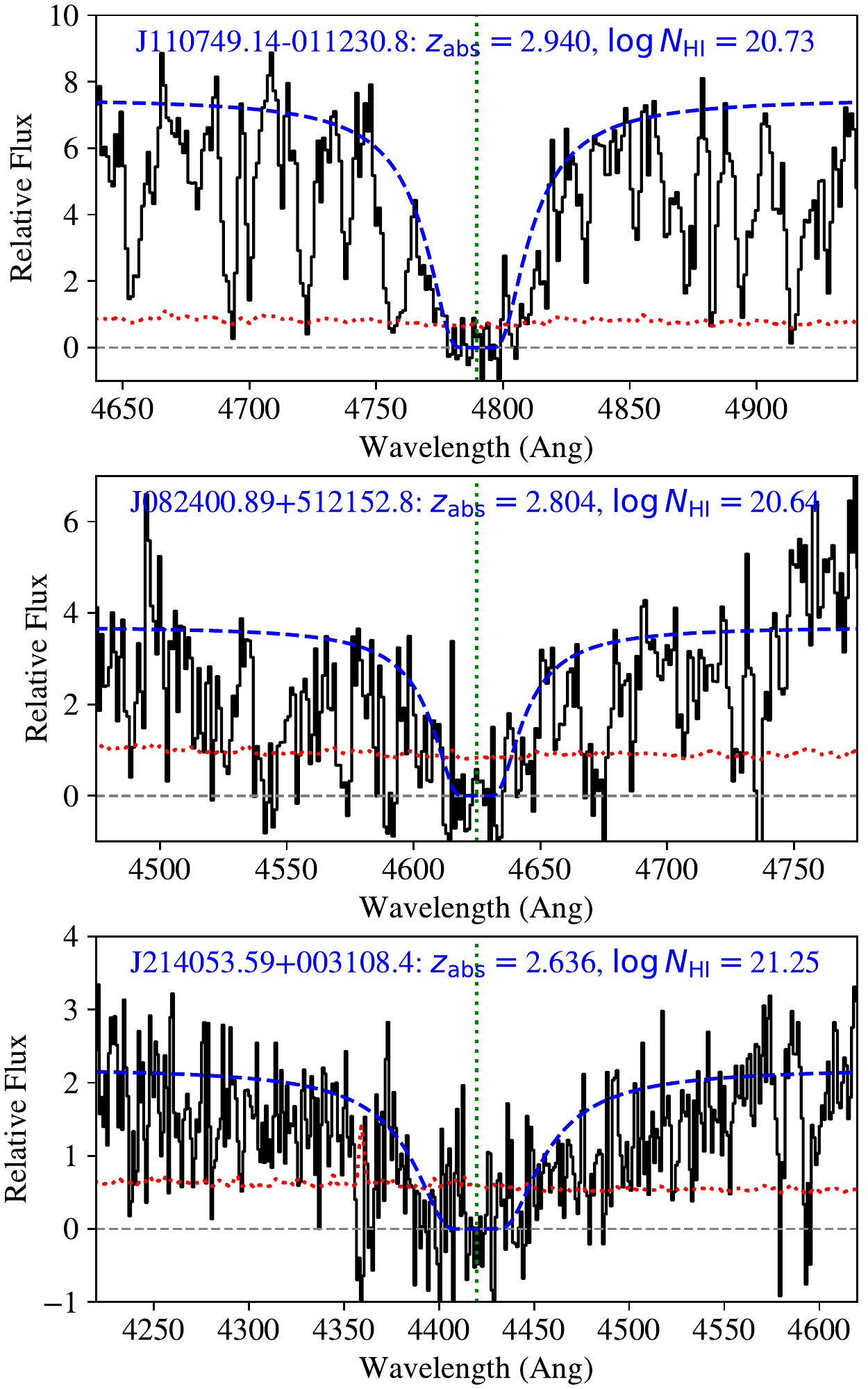}{Three DLAs reported
by the machine learning algorithm that were not previously
reported in prior SDSS DLA surveys.  The top two cases
also show strong, coincident low-ion absorption outside
the \Lya\ forest.  All three quasars were also observed
in the BOSS survey and those data all support their
characterization as DLAs.
}

\section{SDSS-DR7 DLA Catalog}
\label{sec:dr7}

Buoyed by the results from our validation tests, including the comparison to
real world data analyzed by human experts, we applied the CNN model to the
entire SDSS DR7 quasar catalog.  Specifically, we analyzed the 24,740 spectra of
quasars with $\mzem > 2$ from the \cite{SchneiderDR7} catalog provided by
the {\it igmspec} v2 database.  Table~\ref{tab:dr7} provides 
the catalog of DLA candidates reported by the algorithm
including a flag for BAL classification taken from
\cite{shen11} and restricting to systems\footnote{All of the algorithm
outputs are available in our public repository:
https://github.com/davidparks21/qso\_lya\_detection\_pipeline} 
with $\mzabs < \mzem$.  
There are \nsdss\ candidates with $\mzabs < \mzem$
in quasar sightlines not flagged as a BAL,
but we caution that our own visual inspection indicates a non-negligible
fraction of these quasars show associated absorption that mimic DLAs.
If we further restrict to systems with a 
\confp\ parameter $>0.9$, there are \highsdss\
high-confidence DLAs in non-BAL quasar sightlines.

\cite{noterdaeme2009} have performed their own human+algorithm based
survey of the SDSS DR7 quasar spectra.  Matching their sample of 937 DLAs
to ours within $\delta z = 0.015$,  we report 777 of these as DLAs and
45 as SLLS.  For the matched DLAs, we find the \nhi\ values of \cite{noterdaeme2009} are
systematically higher then our algorithm's
by $\approx 0.07$\,dex with a standard deviation
of 0.17\,dex.  We visually inspected the 17 `misses' 
with $\log \mnhi > 20.8$.  Eight are offset by
more than 0.015 in redshift with neither algorithm showing 
systematically superior results.
Seven from the \cite{noterdaeme2009} survey are not DLAs
(e.g.\ the purported system at $\mzabs = 3.563$ 
toward J114317.14+142431.04) or at best
ambiguous.  There is one egregious miss 
(at $\mzabs = 2.88$ towards J133042.52-011927.55) which we then
realized is also absent in the \cite{Prochaska2009} survey.  
Therefore, this spectrum
was part of our DLA-free training set and we expect the
algorithm learned to mis-classify it. 
Lastly, there is one example with
apparent flux in the core of \Lya\ (at $\mzabs = 3.553$ toward
J205509.49-071748.62) which is bonafide and not recovered by our
algorithm (nor \cite{Prochaska2009}).

Altogether we report \newsdss\ DLAs in the DR7 with \confp=1 that
are not listed in \cite{Prochaska2009} nor \cite{noterdaeme2009}
(including over 1,000 at $\mzabs < 3.5$ where blending in the IGM
is less severe).  
We have visually inspected $\approx 100$ of these and find
the majority are in lower S/N data and are highly plausible
DLAs (e.g.\ a high fraction exhibit corresponding, strong
low-ion metal absorption).
Figure~\ref{fig_new_dr7.pdf}
shows several examples with $\mnhi > 10^{20.5} \cm{-2}$. 
Approximately 10\%\ of these high-confidence DLA candidates
are mis-identified associated absorption for
quasars not previously deemed a BAL.
Because \cite{noterdaeme2009} did not publish their survey path
for DLAs, we cannot determine whether any of the additional
DLAs discovered by our algorithm would affect their statistical
results.

\begin{table*}
\centering
\begin{minipage}{170mm} 
\caption{SDSS DR7 DLA CANDIDATES$^a$\label{tab:dr7}}
\begin{tabular}{lcccccccc}
\hline 
RA & DEC & Plate & Fiber & \zabs & \nhi & Conf. & BAL$^b$ 
& Previous?$^c$\\ 
\hline 
145.9092 & -1.0054 & 266 & 124 & 2.322 & 20.41 & 0.47 & 1& 0\\ 
147.0957 & 0.9318 & 266 & 617 & 3.431 & 20.38 & 0.39 & 0& 0\\ 
151.1185 & 0.3071 & 269 & 467 & 2.540 & 20.95 & 0.94 & 0& 3\\ 
151.1185 & 0.3071 & 269 & 467 & 2.685 & 20.95 & 1.00 & 0& 3\\ 
154.6352 & 0.2434 & 271 & 500 & 2.606 & 20.48 & 0.82 & 0& 0\\ 
154.6352 & 0.2434 & 271 & 500 & 3.714 & 20.43 & 1.00 & 0& 0\\ 
\hline 
\end{tabular} 
\end{minipage} 
{$^a$}Restricted to systems with $\mzabs < \mzem$.\\ 
{$^b$}Quasar is reported to exhibit BAL features by \cite{shen11} (1=True).  We caution that additional BAL features exist in the purported non-BAL quasars.\\ 
{$^c$}DLA is new (0) or is also reported by N09 (1), PW09 (2), or both (3).\\ 
\end{table*}

\section{BOSS-DR12 DLA Catalog}
\label{section:dr12}

A principal motivation of this work was to apply a machine learning
algorithm to the complete DR12 dataset.  Using the same model 
derived from the SDSS DR5 training set and applied to the SDSS DR7
quasar spectra, we derive DLAs from the BOSS DR12 dataset provided
in {\it igmspec}, v2.  Specifically, we analyze the 201,776 spectra
for quasars with $\mzem > 1.95$ including sources with BALs.

The model reports \nboss\ 
systems with $\log \mnhi \ge 20.3$,
$\mzabs > 2$, and $\mzabs < \mzem$ 
in the 174,691  
sightlines not flagged as BALs in the BOSS
quasar catalog (Table~\ref{tab:dr12}). 
A 2D histogram of the \nhi\ and \zabs\ values 
is shown in Figure~\ref{fig_ml_boss_dlas.pdf} for the subset
(\highboss) with high confidence (\confp~$>0.9$). 
This relatively pure sample constitutes a several-fold
increase over any previously published catalog.

\imgsinglecol{}{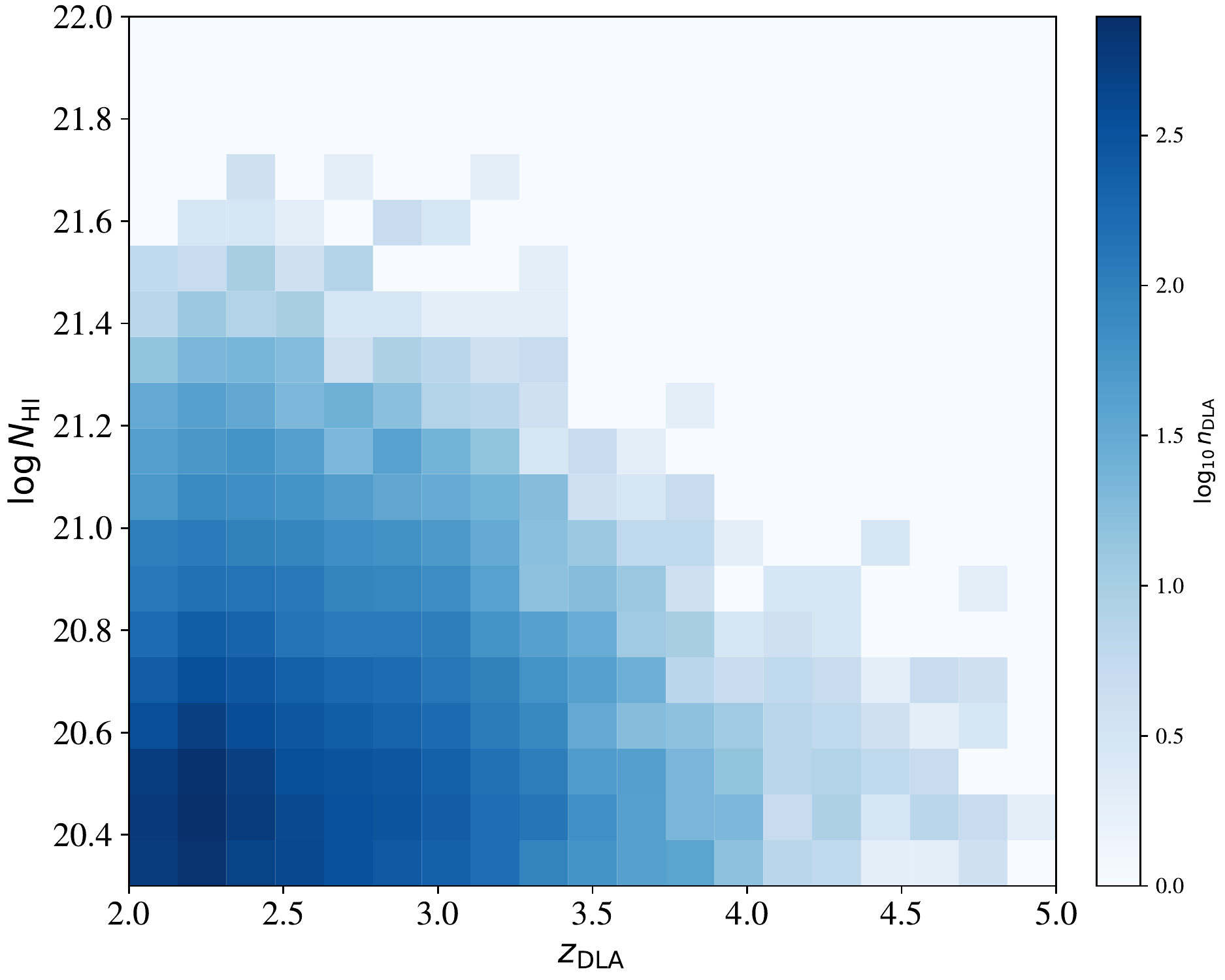}{2D Histogram of the 
\zabs, \nhi\ distribution of DLAs reported by the CNN algorithm
for the set of BOSS DR12 quasars.  Sightlines are restricted to
those without strong BAL features.  The DLAs are restricted to
$\mzabs > 2.$, $\mzabs < \mzem$, and \confp~$>0.9$.  There are 
a total of 18,959 DLAs satisfying these cuts.
}

As described in the Introduction, \cite{Garnett2016} have developed their
own `big data' algorithm to process BOSS DR12 quasar spectra
to assess the likelihood of intervening DLAs.  Matching to
$\delta z = 0.015$ tolerance, we find that 
\bossmatch\
of our pure sample are tabulated in \cite{Garnett2016} with
nearly all occurring in a sightline reported by those authors
to be highly confident to contain a DLA (pDLAD~$>0.9$).
Approximately 40\%\ of the 8,388 non-matched DLAs occur blueward of \Lyb\
in the quasar rest-frame, which was outside the catalog
provided by \cite{Garnett2016}.  
Restricting to quasar rest-frame wavelengths
$\lambda_{\rm rest} > 1030$\AA\ and to DLAs with $\log \mnhi > 21$
(which one expects to be sufficiently strong to detect in any
algorithm), there are 71 systems not matched.  We have inspected
a subset to verify few (if any) are egregious errors in our algorithm.
Most are either in sightlines with 2 or more DLAs (also not classified 
by \cite{Garnett2016}) or proximate to the quasar.
Figure~\ref{fig_boss_hist.pdf} compares the absorption redshifts
and \nhi\ values between the matched DLAs from the two algorithms.
We measure small median offsets of $\Delta z = -0.00035$ and 
$\Delta \mnhi = -0.038$ and standard deviations
of 0.002 and 0.16\,dex respectively.

\imgsinglecol{}{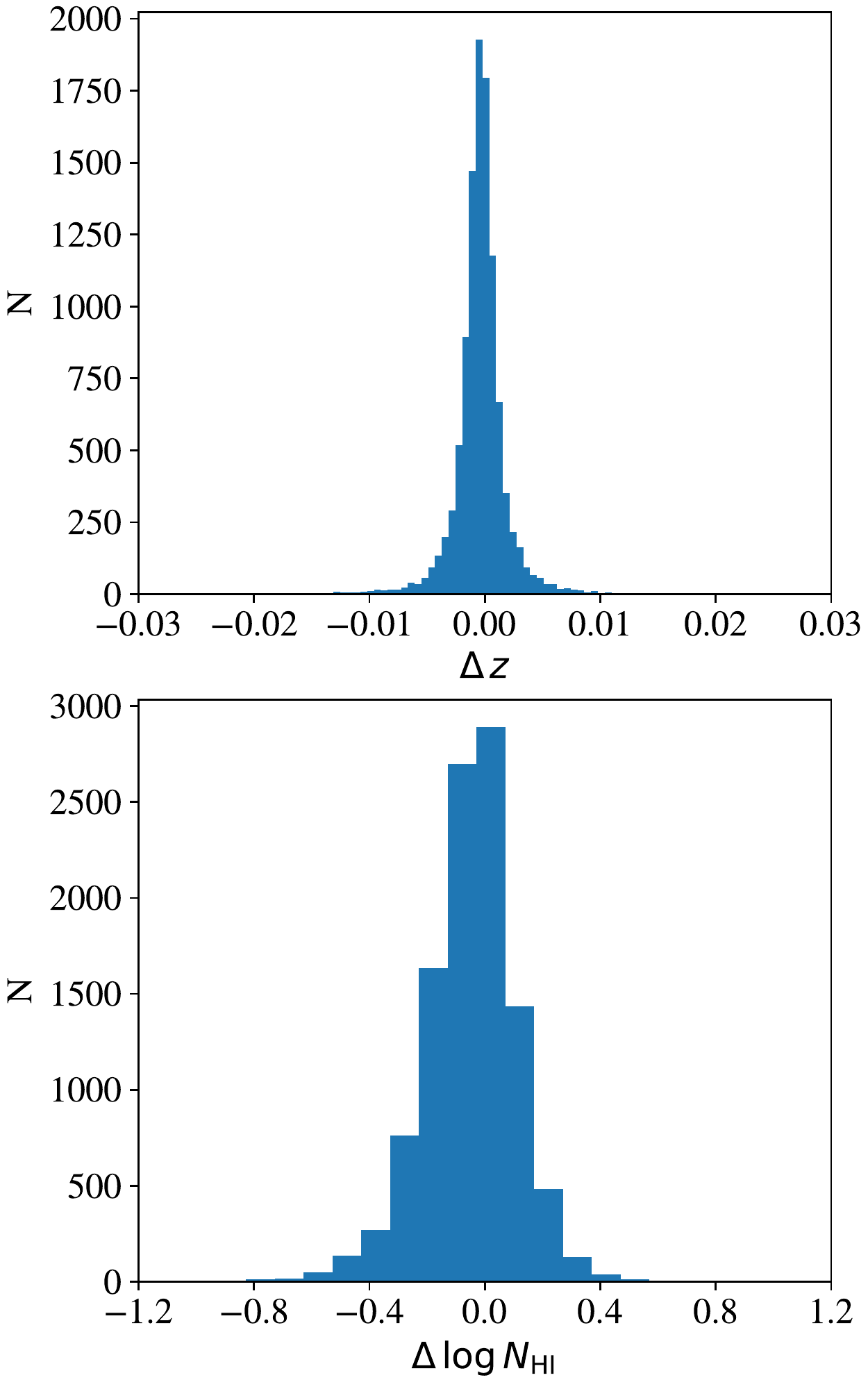}{Histograms showing the
differences for (top) \zabs\ and (bottom) \nhi\ between DLAs
recovered by our deep learning algorithm matched to systems
predicted by Garnett et al.\ (2016).   For these \bossmatch\
matched DLAs, the estimated properties are in excellent agreement.
}

Reversing the matching process, i.e.\ comparing the results 
of \cite{Garnett2016} to our CNN 
algorithm, we find that $\approx 2,000$ of the DLAs in high probability
sightlines are not matched to our algorithm.  Many of these have
\nhi\ close to the DLA threshold but 408 have $\log \mnhi > 21$.
We have examined $\approx 100$ of these cases by inspecting the 
BOSS DR12 spectra at the purported DLA position and also searched
for associated metal-line absorption.
There is a small, subset of cases where the `best' model 
from \cite{Garnett2016} is
very likely in error (e.g.\ bad data, pairs of DLAs,
significant flux in the DLA core; Figure~\ref{fig_g16_junk.pdf}).
The majority, however, correspond to low S/N spectra which exhibit
a broad depression in flux at the putative DLA which may well
result from a high \nhi\ system.  Our algorithm, which was not
extensively trained on very low S/N spectra,
did not classify these cases as DLAs.
There is an additional set of high-\nhi\ misses that 
differ in \zabs\ by more than 0.015 and less than 0.05.
A search for corresponding metal-line absorption near \zabs\ 
shows cases where the model from one or the other
algorithm is preferred and also cases where 
both are significantly offset.
Lastly, there are a few cases (e.g.\ at $\mzabs=3.626$
toward J150731.89+435429.7) where the \cite{Garnett2016} algorithm has
identified a very strong DLA in low S/N spectra that
was missed by the CNN.

\begin{table*}
\centering
\begin{minipage}{170mm} 
\caption{BOSS DR12 DLA CANDIDATES$^a$\label{tab:dr12}}
\begin{tabular}{lcccccccc}
\hline 
RA & DEC & Plate & Fiber & \zabs & \nhi & Conf. & BAL$^b$ 
& G16$^c$?\\ 
\hline 
8.4806 & -0.0140 & 3586 & 158 & 2.098 & 20.45 & 0.84 & 1& 0\\ 
7.7251 & -0.0462 & 3586 & 320 & 2.131 & 20.56 & 0.99 & 0& 0\\ 
7.8980 & 0.7510 & 3586 & 728 & 2.125 & 20.31 & 0.45 & 1& 0\\ 
8.3262 & 0.2714 & 3586 & 816 & 2.478 & 20.30 & 0.78 & 0& 0\\ 
9.8449 & -0.7361 & 3587 & 100 & 2.163 & 20.63 & 1.00 & 0& 1\\ 
9.7215 & -1.0555 & 3587 & 130 & 2.150 & 20.71 & 0.21 & 0& 0\\ 
\hline 
\end{tabular} 
\end{minipage} 
{$^a$}Restricted to systems with $\mzabs < \mzem$ and $\mzabs > 2$.\\ 
{$^b$}Quasar is reported to exhibit BAL features by the BOSS survey.\\ 
{$^c$}DLA is new (0) or reported by G16 (1).\\ 
\end{table*}

\imgsinglecol{}{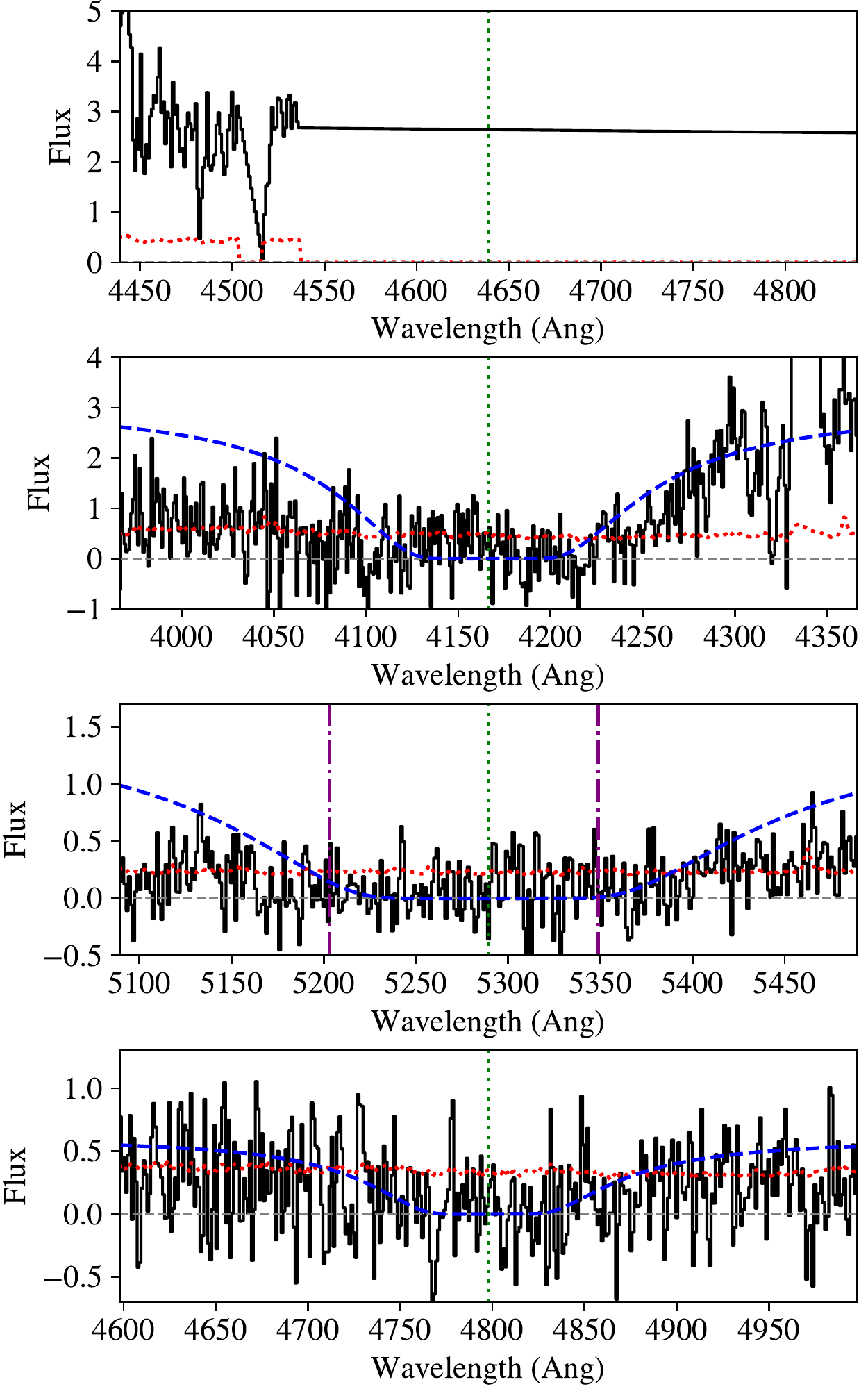}{Example BOSS DR12
spectra at the location of several, high \nhi\ DLAs 
reported in Garnett et al.\ (2016) 
which appear to be in error.   The top
panel is a case of missing data, the 2nd panel has
significant flux in the purported DLA core (a DLA
may be present at higher \zabs), the 3rd panel shows
an example where there are two DLAs (confirmed by
metal-line absorption), and the bottom case 
is a low S/N example with apparent residual flux.
}





\section{Concluding Remarks}
\label{section:conclude}

In this manuscript, we have applied deep learning techniques
on astronomical spectra to classify and characterize the
distinct spectral features known as damped \Lya\ systems (DLAs).
Our fully-automated algorithm efficiently and precisely
identifies DLAs and estimates their absorption redshifts
and HI log column densities, with high efficacy.  The
results, as best we can assess, are as reliable and robust
as any previous human-generated analysis.  The algorithm
struggles in several domains (e.g.\ overlapping and/or very
high \nhi\ DLAs) and frequently misidentifies BALs as DLAs,
but we believe that these issues can be addressed by future,
improved models.

Our results highlight the potential of CNN architectures
for any number of applications with deep learning to 
astronomical spectra.  These could include source classification,
redshift estimation, and the characterization of a wide
variety of absorption and emission features.
As when one applies deep learning techniques to images,
we believe the primary hurdle to success is to generate 
sufficiently large and realistic training sets to span
the dataset that cover (hard-to-predict) edge cases.
For full application to quantitative astrophysical studies,
however, one must also develop techniques to assess both
statistical and systematic uncertainties.
The development of such algorithms for astronomical research
is an essential step in the still emerging era of massive datasets
of astronomical spectra.

\clearpage

\section*{Acknowledgements}

We acknowledge sustained, valuable input from Marcel
Neeleman.  JXP was partially supported by NSF grant
AST-1412981. SD was partially supported by NSF grant 
OAC-1541270.  Our code make use of algorithms contained
within the {\it astropy} (http://www.astropy.org/),
{\it linetools} (https://github.com/linetools/linetools)
and {\it pyigm} (https://github.com/pyigm/pyigm)
packages.



\section{Appendix}
\label{section:appendix}

\subsection{DR12 Catalog Format}
\label{-section:dr12_catalog}

DLA predictions for the Sloan DR12 catalog are provided in two formats: a JSON, containing a hierarchy of Sightline $\rightarrow$ DLA(s), and a CSV format in which multiple DLAs found on a single sightline constitute separate rows in the CSV.

The JSON format, as seen in this section, is an array of sightline entries, one for each sightline in DR12. Each sightline entry is an object/map containing: plate, fiber, ra, dec, z\_qso, num\_dlas, and dlas. Described in table \ref{-table:catalog}.

The dla field contains an array of DLA objects/maps each containing: std\_column\_density, column\_density, spectrum, rest, dla\_confidence, type, z\_dla. Described in table \ref{-table:catalog}.

The CSV contains the same properties listed as a flattened list.
\begin{verbatim}
{
 "mjd": 55181, 
 "plate": 3586, 
 "fiber": 126, 
 "dlas": [
  {
   "std_column_density": 0.17546822130680084, 
   "column_density": 20.314374892580844, 
   "spectrum": 3838.839036881458, 
   "rest": 1162.856689453125, 
   "dla_confidence": 0.3, 
   "type": "DLA", 
   "z_dla": 2.1577969653618645
  }, 
  {
   "std_column_density": 0.14796611666679382, 
   "column_density": 20.914009651215395, 
   "spectrum": 4276.612603011075, 
   "rest": 1295.466552734375, 
   "dla_confidence": 0.41250000000000003, 
   "type": "DLA", 
   "z_dla": 2.5179058486357935
  }
 ], 
 "ra": 8.658418078168893, 
 "num_dlas": 2, 
 "dec": -0.9662403378785263, 
 "z_qso": 2.3012142181396484
}, 
\end{verbatim}



\begin{table*}
\centering
\begin{minipage}{170mm} 
\caption{DR12 JSON and CSV catalog format description}
\begin{tabular}{ll}
\hline 

Property name & Description \\ \hline\hline

plate & Plate sightline identifier for DR12.\\ 

mjd & MJD sightline identifier for DR12. \\ 

fiber & Fiber sightline identifier for DR12. \\ 

ra & RA location of sightline, right ascension. \\ 

dec & DEC location of sightline, declination. \\ 

z\_qso & Quasar redshift as published in DR12. \\ 

num\_dlas & Number of DLAs identified in this sightline. \\&Matches the length of the "dlas" array in the JSON catalog, \\&or indicates how many duplicate sightline entries exist in the CSV. \\ 

dlas & Only in JSON results, contains the list of DLA objects, \\&each of which represents a DLA identified in the sightline. \\ 

dlas $\rightarrow$ column\_density & Log column density ($N_{HI}$) prediction made by the model. \\ 

dlas $\rightarrow$ std\_column\_density & Standard deviation of all log column density estimates made by the model. \\&The model computes column density multiple times over the sliding window model, \\&a high standard deviation would be an indicator that the model was not able to \\&confidently predict the column density. \\ 

dlas $\rightarrow$ spectrum & The location of the DLA in wavelength (Angstroms). \\ 

dlas $\rightarrow$ rest & The location of the DLA in the rest frame (Angstroms). \\ 

dlas $\rightarrow$ z\_dla & The location of the DLA in sightline redshift. \\ 

dlas $\rightarrow$ dla\_confidence & A non statistical measure of confidence $(0,1)$ based on how tightly the model \\&predicted the location of the DLA over the sliding window. \\ 

dlas $\rightarrow$ type & In the published catalog this will always be "DLA", the model can also predict \\&sub-dla's and Ly-$\beta$ absorbers which was not analyzed carefully for accuracy. \\ 

\hline 
\label{-table:catalog}
\end{tabular} 
\end{minipage} 
\end{table*} 


\begin{table*}
\centering
\begin{minipage}{170mm} 
\caption{All hyperparameters used in neural network, these hyperparameters were optimized to produce the best results on held out validation data.}
\begin{tabular}{lcl}
\hline 

Hyperparameter & Value & Description \\
\hline\hline

Learning Rate & 0.0005 to & 
The network is initially trained with a learning rate of 0.0005 \\
& 0.00002 & and the learning rate is manually tuned lower over the training period. \\

Batch Size & 700 & A training batch is computed and averaged for each update. \\

L2 Regularization Penalty & 0.005 & 
L2 regularization penalizes large weights in the network, akin to \\
&&ridge regression. \\

Dropout Keep Probability & 0.98 & 
Dropout regularization was applied, though a keep probability of 98\% \\
&&was found to yield the best results. \\

Convolutional layer 1 filters & 100 &  \\
Convolutional layer 2 filters & 96 & Number of filters for each convolutional layer. \\
Convolutional layer 3 filters & 96 &  \\

Convolutional layer 1 kernel & 32 &  \\
Convolutional layer 2 kernel & 16 & Kernel size for each convolutional layer.\\
Convolutional layer 3 kernel & 16 &  \\

Convolutional layer 1 stride & 3 &  \\
Convolutional layer 2 stride & 1 & Stride along the first \& only dimension of the 1D dataset \\
Convolutional layer 3 stride & 1 &  \\

Pooling layer 1 kernel & 7 &  \\
Pooling layer 2 kernel & 6 & Pooling layers kernel in the first \& only dimension of the 1D dataset. \\
Pooling layer 3 kernel & 6 &  \\

Pooling layer 1 stride & 4 &  \\
Pooling layer 2 stride & 4 & Pooling layer stride in the first \& only dimension of the 1D dataset. \\
Pooling layer 3 stride & 4 &  \\

Shared fully connected layer \# neurons & 350 & 
Size of the first fully connected layer, shared by all outputs. \\

Classification layer \# neurons & 200 & Size of the fully connected layer for classification output. \\
Localization layer \# neurons & 350 & Size of the fully connected layer for localization output. \\
Column density layer \# neurons & 150 & Size of the fully connected layer for column density. \\


\hline 
\label{-table:hyperparameters}
\end{tabular} 
\end{minipage} 
\end{table*} 


\end{document}